\documentclass[11pt,prd,showpacs,superscriptaddress,amsmath,nofootinbib,floatfix]{revtex4}

\usepackage{epsfig}
\usepackage{graphicx}
\usepackage{axodraw}
\usepackage{cancel}
\usepackage{color}
\usepackage{psfrag}
\usepackage{longtable}
\def\axopath{}

\newcommand{\nc}{\newcommand}
\def\theequation{\thesection.\arabic{equation}}
\newcommand{\newsection}[1]{\section{#1}\setcounter{equation}{0}}

\def\nn{\nonumber\\}

\newcommand{\f}{\frac}
\newcommand{\be }{\begin{equation}}   \nc{\ee }{\end{equation}}
\newcommand{\bea}{\begin{eqnarray}}   \nc{\eea}{\end{eqnarray}}
\newcommand{\baa}{\begin{array}}      \nc{\eaa}{\end{array}}
\newcommand{\bit}{\begin{itemize}}    \nc{\eit}{\end{itemize}}
\newcommand{\ben}{\begin{enumerate}}  \nc{\een}{\end{enumerate}}
\newcommand{\eps}{\varepsilon}
\newcommand{\vp}{\varphi}

\newcommand{\Db}{\mbox{${\raisebox{2mm}{\boldmath ${}^\leftarrow$}\hspace{-4mm} D}$}}

\newcommand{\vpj }{\mbox{${\vp^\dag i\,\raisebox{2mm}{\boldmath ${}^\leftrightarrow$}\hspace{-4mm} D_\mu\,\vp}$}}
\newcommand{\vpjt}{\mbox{${\vp^\dag i\,\raisebox{2mm}{\boldmath ${}^\leftrightarrow$}\hspace{-4mm} D_\mu^{\,I}\,\vp}$}}
\def\ocal{{\cal O}}
\def\lcal{{\cal L}}
\def\p{\partial}

\newcommand{\eq}[1]{Eq.~(\ref{#1})}
\begin{document}
\begin{titlepage}
\begin{flushright}
\begin{tabular}{l}
{IFT-10/2013}\\[15mm]
\end{tabular} 
\end{flushright}
\begin{center}
\setlength {\baselineskip}{0.3in} {\bf\Large Lepton Flavor Violation
  in the Standard Model with general Dimension-Six Operators}\\[2cm]
\setlength {\baselineskip}{0.2in} {\large
  A.~Crivellin$^{a,}$\footnote{e-mail: crivellin@itp.unibe.ch},
  S.~Najjari$^{b,}$\footnote{e-mail: saereh.najjari@fuw.edu.pl} and
  J.~Rosiek$^{b,}$\footnote{e-mail: janusz.rosiek@fuw.edu.pl}}\\[5mm]
$^a$~{\it Albert Einstein Center for Fundamental Physics,\\ Institute
  for Theoretical Physics, University of Bern.}\\[5mm]
$^b$~{\it Institute of Theoretical Physics, Department of Physics,
  University of Warsaw.}\\[10mm]
{\bf Abstract}\\[5mm]
\end{center}
\setlength{\baselineskip}{0.2in} 

We study lepton flavor observables in the Standard Model (SM) extended
with all dimension-$6$ operators which are invariant under the SM
gauge group.  We calculate the complete one-loop predictions to the
radiative lepton decays $\mu\to e\gamma$, $\tau \to \mu \gamma$ and
$\tau\to e\gamma$ as well as to the closely related anomalous magnetic
moments and electric dipole moments of charged leptons, taking into
account all dimension-$6$ operators which can generate lepton flavor
violation.  Also the 3-body flavor violating charged lepton decays
$\tau^\pm \to \mu^\pm \mu^+ \mu^-$, $\tau^\pm\to e^\pm e^+ e^- $,
$\tau^\pm \to e^\pm \mu^+ \mu^- $, $\tau^\pm \to \mu^\pm e^+ e^- $,
$\tau^\pm \to e^\mp \mu^\pm \mu^\pm$, $\tau^\pm \to \mu^\mp e^\pm
e^\pm $ and $\mu^\pm \to e^\pm e^+ e^-$ and the $Z^0$ decays
$Z^0\to\ell_i^+\ell_j^-$ are considered, taking into account all
tree-level contributions.  \\[2cm]

\end{titlepage}

\newsection{Introduction \label{sec:intro}}

The Standard Model (SM) of strong and electroweak interactions has
been successfully tested to a great precision~\cite{PDG}.
Nevertheless, it is commonly accepted that it constitutes only an
effective theory which is valid up an energy scale $\Lambda$ where new
physics (NP) enters and additional dynamic degrees of freedom become
important. A renormalizable quantum field theory valid above this
scale should satisfy the following requirements:
\bit
\item[({\it i})] Its gauge group must contain the SM gauge
  group~$SU(3)_C\times SU(2)_L\times U(1)_Y$.
\item[({\it ii})] All SM degrees of freedom should be incorporated
  either as fundamental or as composite fields.
\item[({\it iii})] At low-energies it should reduce to the SM provided
  no undiscovered weakly coupled {\em light} particles exist (like
  axions or sterile neutrinos).
\eit

In most theories of physics beyond the SM that have been considered,
the SM is recovered at low energies via the decoupling of the heavy
particles with masses of the order of $\Lambda\gg M_Z$.  That such a
decoupling at the perturbative level is possible in a renormalization
quantum field theory is guaranteed by the Appelquist-Carazzone
decoupling theorem~\cite{Appelquist:1974tg}.  This leads to the
appearance of higher-dimensional operators which are suppressed by
powers of $\Lambda$ and are added to the SM Lagrangian:
\be 
\label{Leff} 
\lcal_{\mathrm SM} = \lcal_{\mathrm SM}^{(4)} + \f{1}{\Lambda }
\sum_{k} C_k^{(5)} Q_k^{(5)} + \f{1}{\Lambda^2} \sum_{k} C_k^{(6)}
Q_k^{(6)} + \ocal\left(\f{1}{\Lambda^3}\right)\,.
\ee
Here $\lcal_{\mathrm SM}^{(4)}$ is the usual renormalizable part of
the SM Lagrangian which contains dimension-$2$ and dimension-$4$
operators only.  $Q_k^{(5)}$ is the Weinberg operator giving rise to
neutrino masses, $Q_k^{(6)}$ denote dimension-$6$ operators, and
$C_k^{(n)}$ stand for the corresponding dimensionless coupling
constants, i.e. the Wilson coefficients.

Even if the ultimate theory of NP at some high energy scale is not a
quantum field theory, at low energies the effective theory still
reduces to a quantum field theory~\cite{Weinberg:1995mt} and it is
possible to parametrize its effects at the electroweak scale in terms
of these operators and the associated Wilson coefficients.  Thus, one
can search for NP in a model independent way by studying the SM
extended with gauge invariant effective higher dimensional operators.
Later, once a specific model is chosen, the Wilson coefficients can be
calculated as a function of model parameters by matching the model of
NP under consideration on the SM extended with such higher dimensional
operators and one can calculate bounds on the specific model as well.

Flavor observables, especially flavor changing neutral current (FCNC)
processes are an excellent probe of new physics since they are
suppressed in the SM and therefore sensitive even to small NP
contributions.  This also means that these processes can stringently
constrain the Wilson coefficients of the dimension-$6$ operators
induced by NP.

Especially the search for lepton flavor violation (LFV) is very
promising since in the SM (extended with massive neutrinos) all flavor
violating effects in the charged lepton sector are proportional to the
very small neutrino masses - e.g. the decay rates of heavy charged
leptons into lighter ones are suppressed by the ratio $m_\nu^2/M_W^2$
and thus are by far too small to be measurable in any foreseeable
experiment.  This in turn means that any observation of LFV would
prove the existence of physics beyond the SM.  In addition, LFV
processes have the advantage of being ``theoretically clean'',
i.e. they can be computed precisely without problems with
non-perturbative QCD effects affecting similar observables in the
quark sector.

Also the current experimental situation and prospects for the search
for charged lepton flavor violation are very promising.  In
Tables~\ref{table:RLFVdecays} and~\ref{table:llllEXP} we list the
experimental bounds on the radiative lepton decays $\ell_i\to
\ell_f\gamma$ and on the three-body lepton decays
$\ell_i\to\ell_j\ell_k\ell_l$, respectively.  Especially the limits on
$\mu \to e$ transitions are very stringent due to constraints from the
MEG and SINDRUM collaborations at the PSI and will be even further
improved in the future: MEG can measure ${\rm Br}[\mu \to e \gamma]$
down to $6\times 10^{-14}$ and a MEG upgrade \cite{Baldini:2013ke}
could increase the sensitivity by another order of magnitude.
Furthermore, the electric dipole moments (EDM) and the anomalous
magnetic moments of charged leptons are theoretically closely related
to $\ell_i\to \ell_f\gamma$ transitions and also here the experimental
accuracies are very good, leading to strong upper bounds for the EDMs
(see Table~\ref{tab:EDMs}). In addition, there is a longstanding
discrepancy between the SM prediction and the measurement of the
anomalous magnetic moment of the muon, which might be a hint for
physics beyond the SM.

\begin{table}[htdp]
\begin{minipage}{2in}
\centering \vspace{0.8cm} 
\renewcommand{\arraystretch}{1.2}
  \begin{tabular}{@{}|c|c|}
\hline 
Process & Experimental bound \\
\hline \hline
$\mathrm{Br}\left[ \tau \to \mu \gamma \right] $ & ~$4.4 \times
10^{-8}$ ~\cite{Aubert:2009ag,Hayasaka:2007vc} \\
\hline 
$\mathrm{Br}\left[ \tau \to e \gamma \right] $ & ~$3.3 \times
10^{-8}$~ \cite{Aubert:2009ag} \\
\hline
$ \mathrm{Br}\left[ \mu \to e \gamma \right] $ & ~$5.7 \times
10^{-13}$~\cite{Adam:2013mnn} \\
\hline \hline
\end{tabular}
 \end{minipage} 
 \caption{Experimental upper limits on the branching ratios of the
   radiative lepton decays.}
\label{table:RLFVdecays}
\end{table}

\begin{table}[htdp]
\begin{minipage}{2in}
\centering \vspace{0.8cm}
\renewcommand{\arraystretch}{1.2}
  \begin{tabular}{@{}|c|c|}
 \hline 
Process & Experimental bound \\ \hline \hline
$\mathrm{Br}\left[\tau^-\to\mu^-\mu^+\mu^-\right] $ & ~$2.1 \times
10^{-8}$ ~\cite{Hayasaka:2010np} \\
\hline 
$\mathrm{Br}\left[\tau^-\to e^-e^+e^-\right] $ & ~$2.7\times 10^{-8}$
~\cite{Hayasaka:2010np} \\
\hline 
$\mathrm{Br}\left[\tau^-\to e^- \mu^+\mu^- \right] $ & ~$2.7 \times
10^{-8}$~ \cite{Hayasaka:2010np} \\
\hline
$\mathrm{Br}\left[\tau^-\to \mu^- e^+\mu^- \right] $ & ~$1.7 \times
10^{-8}$~ \cite{Hayasaka:2010np} \\
\hline $ \mathrm{Br}\left[\mu^-\to e^-e^+e^- \right]$ & ~$1.0 \times
10^{-12}$~\cite{Bellgardt:1987du} \\
\hline \hline
\end{tabular}
\end{minipage} 
\caption{Experimental upper limits on the branching ratios of the
  three body charged lepton decays.}
\label{table:llllEXP}
\end{table} 

\begin{table}[htdp]
\centering \vspace{0.8cm}
\renewcommand{\arraystretch}{1.2}
\begin{tabular}{|c| c| c| c| c|}
\hline \hline
EDM & ${\left|d_{e}\right|}$ & ${|d_{\mu}|}$ & ${d_{\tau}}$ \\
\hline \hline
Bound $\left[\rm e \, cm\right]$ & $8.7 \times 10^{-29}$
\cite{Baron:2013eja} & $1.9 \times 10^{-19} $ \cite{Bennett:2008dy} &
$ \left[ -2.5, \,0.8 \right] \times 10^{-17}$ \cite{Inami:2002ah} \\
\hline \hline
\end{tabular}
\caption{Experimental upper bounds (or allowed range for ${d_{\tau}}$)
  on electric dipole moments of the charged leptons.}
\label{tab:EDMs}
\end{table}

Lepton flavor violating processes have been studied in great detail in
many specific extensions of the SM.  For example in the MSSM
non-vanishing decay widths for LFV processes are generated by flavor
non-diagonal SUSY breaking terms \cite{Borzumati:1986qx,
  Brignole:2004ah, Paradisi:2005fk, Altmannshofer:2009ne,
  Girrbach:2009uy}.  Also extending the MSSM with right-handed
neutrinos by the seesaw mechanism \cite{Minkowski:1977sc} gives rise
to LFV \cite{Ilakovac:1994kj,Hisano:1995cp, Hisano:2001qz,
  Babu:2002et, Masiero:2004js, Dedes:2007ef, Antusch:2006vw,
  Ilakovac:2012sh,Ilakovac:2013wfa}, as well as allowing for R-parity
violation \cite{deGouvea:2000cf, Abada:2001zh, Dedes:2006ni}.  Other
models like the littlest Higgs Model with T-Parity
\cite{Blanke:2007db}, two-Higgs-doublet models with generic flavor
structures \cite{Kanemura:2004cn, Kanemura:2005hr, Paradisi:2005tk,
  Crivellin:2013wna} or models with an extended fermion sector
\cite{Buras:2011ph} have sources of lepton flavor violation, too.  In
order to make models of New Physics consistent with the
non-observation of LFV processes in Nature, the assumption of Minimal
Flavor Violation \cite{MFV} has been extended to the lepton sector
(see e.g.  \cite{Cirigliano:2005ck, Nikolidakis:2007fc}).  Flavor
changing $\tau$ decays have been studied in
Ref.~\cite{Dassinger:2007ru} in a model independent way taking into
account a (reducible) set of four-lepton operators and the magnetic
lepton operators. However, a detailed model independent analysis with
all gauge invariant operators is still pending\footnote{For a model
  independent analysis for the Higgs sector of the SM see
  Ref.~\cite{AguilarSaavedra:2008zc, Contino:2013kra} and for
  anomalous top couplings Ref.~\cite{Grzadkowski:2003tf,
    AguilarSaavedra:2008zc}.}.

In this article we perform such a model independent analysis by
considering the SM extended with all dimension-$6$ operators giving
rise to lepton flavor violation which are invariant under the SM gauge
group.  We study the radiative lepton decays $\ell_i \to \ell_f
\gamma$ and three-body charged lepton decays
$\ell_i\to\ell_j\ell_k\ell_l$, as well as the anomalous magnetic
moments and EDMs of charged leptons and the flavor violating $Z^0\to
\ell_i^-\ell_j^+$ decays.

It is worth noting that analyzing the LFV processes using the
gauge-invariant basis of dimension-$6$ operators automatically assures
that the final results are also gauge invariant and contain all
relevant contributions. Otherwise, one risks including just subset of
diagrams contributing to a given process.  For example it is quite
common in the literature to calculate in a model of NP only the
effective flavor changing $Z^0$-boson coupling to charged leptons and
neglect the corrections to $W$ couplings, as the latter do not
contribute at the tree-level to neutral current processes.  However,
both $Z^0$ and $W$ (and also Goldstone boson) couplings come from the
same set of gauge-invariant higher-order operators, and are thus of
the same size. In fact, (as our calculation shows explicitly) their
contributions at least to some processes, like
e.g. $\ell_i\to\ell_f\gamma$, are equally important and should be
always considered together.

The outline of this article is as follows: after recalling the
relevant dimension-$6$ operators in the next Section we will consider
radiative lepton decays in Sec.~\ref{sec:llg} (including the related
anomalous magnetic moments and electric dipole moments of charged
leptons), three-body charged lepton decays in Sec.~\ref{sec:llll} and
the flavor changing $Z^0$ decays, $Z^0\to \ell_i^-\ell_j^+$, in
Sec.~\ref{sec:zll}.  We calculate the full one-loop predictions for
the $\ell_i \to \ell_f \gamma$ decays and all tree-level contributions
for $\ell_i\to\ell_j\ell_k\ell_l$ decays in terms of the Wilson
coefficients of the dimension-$6$ operators.  Sec.~\ref{sec:numerics}
deals with the numerical evaluation of our results and finally we
conclude in Sec.~\ref{sec:concl}.  An Appendix summarizes the Feynman
rules arising from the dimension-$6$ operators after electroweak
symmetry breaking and the additional form-factors for
$\ell_i\to\ell_f\gamma^*$ amplitude for the case of an off-shell
photon.

\newsection{The lepton flavor violating operators of
  dimension-$6$\label{sec:oper}}

The complete (but still reducible) list of independent operators of
dimension-$5$ and dimension-$6$ which can be constructed out of SM
fields and which are invariant under the SM gauge group fields was
first derived in Ref.~\cite{Buchmuller:1985jz}.  In this article we
follow the notation Ref.~\cite{GIMR} where the operator basis of
Ref.~\cite{Buchmuller:1985jz} was reduced to a minimal set.  For
completeness, we list below again the operators relevant for our
discussion.  We use the following indices and symbols:
\begin{itemize}
\item $a,b=1,2$ label the components of the weak isospin doublets.
\item $i,j,k,l$ are flavor indices running from 1 to 3.
\item $L$ and $R$ stand for the chiralities.
\item $\ell_i = \left( \begin{array}{c} \nu_{Li}
  \\ \ell_{Li} \end{array} \right)$ and $q_i = \left( \begin{array}{c}
  u_{Li}\\ d_{Li} \end{array} \right)$ stand for the lepton and the
  quark doublets.
\item $e_i=\ell_{Ri}$, $u_i=u_{Ri}$ and $d_i=d_{Ri}$ are the
  right-handed isospin singlets.
\item $\varphi^a$ is the SM Higgs doublet where $\varphi^2$ is the
  neutral component.
\end{itemize}

\begin{table}[t]
\begin{center}
\begin{tabular}{|c|p{4mm}cp{4mm}cp{4mm}cp{4mm}cp{4mm}cp{4mm}|c|}
\hline
& \multicolumn{11}{|c|}{fermions} & scalars \\[1mm]
\hline
field && $\ell^a_{Li}$ && $e_{Ri}$ && $q^a_{Li}$ && $u_{Ri}$ &&
$d_{Ri}$ && $\vp^a$ \\[1mm]
\hline
hypercharge $Y$ && $-\frac{1}{2}$ && $-1$ && $\frac{1}{6}$ &&
$\frac{2}{3}$ && $-\frac{1}{3}$ && $\frac{1}{2}$ \\[1mm] 
\hline 
\end{tabular}
\end{center}
\caption{Our conventions for the hypercharges of the SM
  fields. \label{tab:matter}}
\end{table}

The hypercharges of the SM fields are summarized in
Table~\ref{tab:matter}.  The sign convention for the covariant
derivatives is
\be 
\label{D-sign} 
\left( D_\mu \ell \right)^{a} = \left( \delta_{ab} \p_\mu +
\frac{1}{2}i g \tau^I_{ab} W^I_\mu + i g^\prime Y_\ell\, \delta_{ab}
B_\mu \right) \ell^{b}\,.
\ee
with $\tau^I$ being the Pauli matrices.  The hermitian derivative
terms are ($\vp^\dag \Db_\mu\vp \equiv (D_\mu\vp)^\dag\vp$):
\be \label{Dfb}
\vpj \equiv i \vp^\dag \left( D_\mu - \Db_\mu \right)
\vp\mbox{\hspace{5mm} and \hspace{5mm}} \vpjt \equiv i \vp^\dag
\left(\tau^I D_\mu - \Db_\mu \tau^I \right) \vp\,.
\ee
The gauge field strength tensors read
\bea W_{\mu\nu}^I &=& \p_\mu W_\nu^I - \p_\nu W_\mu^I - g \eps^{IJK}
W_\mu^J W_\nu^K\,, \\
B_{\mu\nu} &=& \p_\mu B_\nu - \p_\nu B_\mu\,
\,.
\eea

In general, the SM can be extended by higher dimensional operators
starting from dimension-$5$.  However, there is just a single
dimension-$5$ term respecting the SM gauge symmetry which, after
electroweak symmetry breaking, generates neutrino masses and mixing
angles - the Weinberg operator ($C$ is the charge conjugation matrix
and $\eps_{12}=+1$):
\be 
\label{qnunu} 
Q_{\nu\nu} = \eps_{ab} \eps_{cd} \vp^a \vp^c(\ell^b_i)^T C \ell^d_j \,.
\ee
This operator does not contribute directly (other then modifying the
$U_{PMNS}$ matrix) to LFV processes in the charged lepton sector,
consequently we do not consider it in the rest of the paper.

In Table~\ref{tab:dim6} we collect the independent dimension-$6$
operators relevant for our discussion, i.e. all operators which can
contribute to LFV processes in the charged lepton sector at the
tree-level or at the 1-loop level.  We neglect the operators which
could give LFV effects only via the interference with the dimension-4
SM vertices containing the PMNS matrix, since such effects are
suppressed by the small neutrino masses which we assume to be zero.
The names of operators in the left column of each block should be
supplemented with generation indices of the fermion fields whenever
necessary, e.g. $Q_{\ell q}^{(1)} \to Q_{\ell q}^{(1)ijkl}$.  Dirac
and color indices (not displayed) are always contracted within the
brackets.  The same is true for the isospin indices, except for
$Q_{\ell equ}^{(1)}$ and $Q_{\ell equ}^{(3)}$.

\begin{table}[htb] 
\centering
\renewcommand{\arraystretch}{1.5}
\begin{tabular}{||c|c||c|c||c|c||} 
\hline \hline
\multicolumn{2}{||c||}{$\ell\ell\ell\ell$} &
\multicolumn{2}{|c||}{$\ell\ell X\vp$} &
\multicolumn{2}{|c||}{$\ell\ell\vp^2 D$ and $\ell\ell\vp^3$}\\ 
\hline
$Q_{\ell \ell }$ & $(\bar \ell_i \gamma_\mu \ell_j)(\bar
\ell_k\gamma^\mu \ell_l)$ & $Q_{eW}$ & $(\bar \ell_o \sigma^{\mu\nu}
e_j) \tau^I \vp W_{\mu\nu}^I$ & $Q_{\vp \ell }^{(1)}$ & $(\vpj)(\bar
\ell_i\gamma^\mu \ell_j)$\\
$Q_{ee}$ & $(\bar e_i \gamma_\mu e_j)(\bar e_k \gamma^\mu e_l)$ &
$Q_{eB}$ & $(\bar \ell_i \sigma^{\mu\nu} e_j) \vp B_{\mu\nu}$ &
$Q_{\vp \ell}^{(3)}$ & $(\vpjt)(\bar \ell_i \tau^I \gamma^\mu
\ell_j)$\\
$Q_{\ell e}$ & $(\bar \ell_i \gamma_\mu \ell_j)(\bar e_k \gamma^\mu
e_l)$ & & & $Q_{\vp e}$ & $(\vpj)(\bar e_i \gamma^\mu e_j)$\\
& & & & $Q_{e\vp 3}$ & $(\vp^\dag\vp)(\bar\ell_i e_j\vp)$\\
\hline \hline
\multicolumn{6}{||c||}{$\ell\ell qq$} \\
\hline 
$Q_{\ell q}^{(1)}$ & $(\bar \ell_i \gamma_\mu \ell_j)(\bar q_k
\gamma^\mu q_l)$ & $Q_{\ell d}$ & $(\bar \ell_i \gamma_\mu
\ell_j)(\bar d_k \gamma^\mu d_l)$ & $Q_{\ell u}$ & $(\bar \ell_i
\gamma_\mu l_j)(\bar u_k \gamma^\mu u_l)$ \\
$Q_{\ell q}^{(3)}$ & $(\bar \ell _i \gamma_\mu \tau^I \ell _j)(\bar
q_k \gamma^\mu \tau^I q_l)$ & $Q_{ed}$ & $(\bar e_i \gamma_\mu
e_j)(\bar d_k\gamma^\mu d_l)$ & $Q_{eu}$ & $(\bar e_i \gamma_\mu
e_j)(\bar u_k \gamma^\mu u_l)$ \\
$Q_{eq}$ & $(\bar e_i \gamma^\mu e_j)(\bar q_k \gamma_\mu q_l)$ &
$Q_{\ell edq}$ & $(\bar \ell _i^a e_j)(\bar d_k q_l^a)$ & $Q_{\ell
  equ}^{(1)}$ & $(\bar \ell _i^a e_j) \eps_{ab} (\bar q_k^b u_l)$
\\ 
&& &&
$Q_{\ell equ}^{(3)}$ & $(\bar \ell _i^a \sigma_{\mu\nu} e_a) \eps_{ab}
(\bar q_k^b \sigma^{\mu\nu} u_l)$ \\
\hline\hline
\end{tabular}
\caption{Complete list of the dimension-$6$ operators (invariant under
  the SM gauge group) which contribute to the LFV observables under
  consideration at the tree or at the one-loop level.
\label{tab:dim6}}
\end{table}
Note that different flavor index combinations of the 4-lepton
operators can correspond to the same operator (for example
$Q_{\ell\ell}^{ijkl} = Q_{\ell\ell}^{ilkj}= Q_{\ell\ell}^{kjil}=
Q_{\ell\ell}^{klij}$).  For this reason, in the following we will only
consider one of these combinations which avoids the introduction of
combinatorial factors.  This can be achieved by the requirement $i\geq
k$, $j\geq l$ for $Q_{\ell\ell,ee}^{ijkl}$, so that the relevant part
of the Lagrangian can be written as:
\bea
{\cal L} = \frac{1}{\Lambda^2}\sum_{ijkl, i\geq k, j\geq l}
\left(C_{\ell\ell}^{ijkl} Q_{\ell\ell}^{ijkl} + C_{ee}^{ijkl}
Q_{ee}^{ijkl}\right) + \frac{1}{\Lambda^2}\sum_{ijkl} C_{\ell
  e}^{ijkl} Q_{\ell e}^{ijkl}\,.
\eea
Note that for $C_{\ell e}^{ijkl}$ all possible flavor index
permutations correspond to different operators.  Due to the
hermiticity of the Lagrangian we find the additional relations like
$C_{\ell\ell}^{ijkl} = C_{\ell\ell}^{jilk\star}$.  Similar ones hold
for all four-fermion operators.

The dominant contributions to the processes considered in this article
are given by diagrams with flavor changing gauge boson vertices or
contact 4-fermion vertices.  However, to preserve gauge-invariance,
also Goldstone boson exchanges has to be taken into account even if,
with few exceptions of mixed $W^\pm G^\mp$ diagrams, they are
suppressed by additional powers of light lepton masses over $v$, the
Higgs field VEV.  In general, the operators listed in
Table~\ref{tab:dim6} give rise also to flavor violating physical Higgs
boson couplings.  We neglect them in our analysis as they are again of
the higher order in $m_\ell/m_{h^0}$.

The $(\vp^\dag\vp)(\bar\ell_i e_j\vp)$ operator does not contain gauge
boson fields and modifies only Higgs and Goldstone boson couplings,
which in principle could affect our results.  However, it gives also
new ${\cal O}(1/\Lambda^2)$ contribution to the charged lepton mass
matrix:
\begin{equation}
m_{fi}^\ell = \dfrac{v}{\sqrt{2}} Y^\ell_f \delta_{fi} +
\dfrac{v^3}{2\sqrt{2}\Lambda^2} C^{fi}_{e\vp 3}\,.
\end{equation}
The necessary rediagonalization of lepton masses has the effect of
modifying the relation between the Yukawa coupling and the charged
lepton masses (and the PMNS matrix). However, one can see that in the
triple Goldstone boson couplings to leptons still the physical lepton
masses and the physical PMNS matrix enter so the $Q^{fi}_{e\phi 3}$
does not generate flavor violation in these couplings.  The triple
coupling of the physical Higgs boson $h^0$ to charged leptons, as well
as all quadruple and quintuple vertices derived from $Q^{fi}_{e\phi
  3}$ can still be flavor violating. Nonetheless, their contributions
to the processes discussed below vanish or are small due to an
additional suppression of $m_\ell/m_{h^0}$, compared to the dominant
contributions from $Q_{\vp e}$, $Q_{\vp \ell}^{(1)}$ and $Q_{\vp
  \ell}^{(3)}$ operators\footnote{$O^{fi}_{e\vp 3}$ generates
  flavour-changing couplings of the SM-Higgs. The resulting effects
  have been studied in Refs.~\cite{Goudelis:2011un,
    Blankenburg:2012ex, Harnik:2012pb} }.  Thus, we neglect this
operator (and thus the entire $\ell\ell\vp^3$ class) in our analysis,
provided that the rediagonalization of the lepton mass matrix has been
performed.

The operators of the $\ell\ell X\vp$ class (as defined in
Table~\ref{tab:dim6}) can give rise to both radiative lepton decays
and to three-body neutral current lepton decays already at the
tree-level.  The 4-lepton $\ell\ell\ell\ell$ operators and the
operators of the $\ell\ell \vp^2 D$ class can contribute to $\ell_i\to
\ell_j\ell_k\ell_l$ decays at the tree-level and to $\ell_i\to \ell_f
\gamma$ decays at the 1-loop level.  Finally, the operators of the
$\ell\ell qq$ class can contribute to both types of decays only at the
1-loop level.  However, for 3-body decays we are only interested in
the tree-level contributions and concerning the radiative lepton
decays, it turns out that only $Q^{(3)}_{\ell equ}$ gives a non-zero
contribution.

In the Appendix we list the Feynman rules arising from the operators
given in Table~\ref{tab:dim6} which are necessary in order to
calculate the flavor observables discussed in the next Sections.

\newsection{Observables related to the effective lepton-photon
  coupling\label{sec:llg}}

As outlined in the introduction, observables related to effective
lepton-photon coupling: radiative lepton decays (especially $\mu\to
e\gamma$), EDMs of charged leptons and their anomalous magnetic
moments are very sensitive to NP and allow to constrain stringently
the relevant Wilson coefficients.

The general form of the flavor violating photon-lepton vertex can be
written as:
\begin{equation}
V_{\ell \ell\gamma}^{fi\;\mu} =
\frac{i}{\Lambda^2}\left[\gamma^\mu(F_{VL}^{fi} P_L + F_{VR}^{fi} P_R)
  + (F_{SL}^{fi} P_L + F_{SR}^{fi} P_R) q^\mu + (F_{TL}^{fi}
  i\sigma^{\mu\nu} P_L + F_{TR}^{fi} i\sigma^{\mu\nu} P_R)
  q_\nu\right]\,.
\label{eq:genform}
\end{equation}
In this Section we calculate the expressions for the formfactors
in~\eq{eq:genform} necessary to calculate the branching ratio for the
$\ell_i\to \ell_f \gamma$ decays (with $i>f$) at the 1-loop level up
the order $1/\Lambda^2$.  In addition, the obtained results are
directly related to the anomalous magnetic moments and the electric
dipole moments (EDM) of leptons after setting $f=i$.

\subsection{Radiative lepton decays\label{sec:llg1}}

Gauge-invariance requires that $F_{VL}$ and $F_{VR}$ must vanish for
on-shell external particles.  The form-factors $F_{SL}$ and $F_{SR}$
do not contribute to the $\ell_i\to \ell_f \gamma$ decay amplitude and
the branching ratio can be expressed in terms of $F_{TL}^{fi}$ and
$F_{TR}^{fi}$ only:
\begin{equation}
\mathrm{Br} \left[\ell_{i} \to \ell_{f}
  \gamma\right]\,=\,\dfrac{m_{\ell_i}^3}{16\pi\Lambda^4 \,
  \Gamma_{\ell_i}} \left( \left|F^{fi}_{TR} \right|^2 + \left|
F^{fi}_{TL} \right|^2\right ) \,.
\label{Brmuegamma}
\end{equation}
The total decay width of the muon is given by $\Gamma_\mu =
\frac{G_F^2 m_\mu^5}{192\pi^3}$ and for the tau lepton $\Gamma_\tau$
includes the leptonic and hadronic decay channels.

Only the operators $Q_{eW}$ (here $W$ denotes the neutral gauge boson
of the $SU(2)_L$ gauge group) and $Q_{eB}$ can contribute to
$F_{TL,R}^{fi}$ at the tree-level. If their coefficients are
comparable to other Wilson coefficient of the dimension~6 operators,
they dominate the effective photon-lepton vertex, with the
form-factors simply given by ($v=\frac{2M_W}{g_2}$):
\bea
F_{TR}^{fi} = F_{TL}^{if\star} = v\sqrt{2} \left(c_W C_{eB}^{fi} - s_W
C_{eW}^{fi} \right) \equiv v\sqrt{2} C_\gamma^{fi}\,.
\label{eq:phtree}
\eea
However, in a renormalizable theory of NP the operators $Q_{eW}$ and
$Q_{eB}$ can only be generated at the loop-level while other
operators, like the effective four-lepton couplings, can already be
generated at the tree-level.  In some extensions of the SM $C_{eW}$
and $C_{eB}$ may even not be generated at all ~\cite{Einhorn:2013kja}.
Thus, comparable (or even dominant) contributions to the flavor
violating lepton-lepton-photon vertex can come from other
dimension-$6$ operators, which for consistency should be included at
the 1-loop level.  The generic topologies of the diagrams which could
contribute to $\ell_i\to \ell_f \gamma$ at the 1-loop level in the
order $1/\Lambda^2$ and the relevant momenta assignments are shown in
Fig.~\ref{fig:llgen}.

\input{\axopath topology.axo}

The list of all 1-loop diagrams contributing to the effective
lepton-photon vertex is given in Fig.~\ref{fig:llgse} (lepton
self-energy contributions) and Fig.~\ref{fig:llg1pi} (1-particle
irreducible vertex corrections).  The diagrams contributing to
photon-photon and $Z^0$-photon self-energies are the same as in the SM
(with $W$ bosons, charged ghosts, charged Goldstone bosons and charged
fermion as virtual particles).  In our loop calculations we do not
take into account flavor violating photon and $Z^0$ couplings
generated at the one-loop level by the operators $Q_{eW}$ and $Q_{eB}$
because if their coefficients are non-negligible, than already the
tree-level contribution of~\eq{eq:phtree} would dominate the whole
process anyway.

\input{\axopath se.axo} 

\input{\axopath triangle.axo}

Our final 1-loop results for the form-factors $F_{TL}$ and $F_{TR}$
are given in Table~\ref{tab:llgres}.  We group them into subsets;
within these subsets the vector form-factors $F_{VL}$ and $F_{VR}$
vanish separately in the on-shell limit.  We kept only the leading
term in $1/\Lambda^2$ and we expand all diagrams involving $Z^0$ and
$W$ bosons (or the associated Goldstone bosons) in the charged lepton
masses, keeping only the leading terms in $m_{\ell}/m_W$,
$m_{\ell}/m_Z$.  For this expansion we used two independent approaches
for calculating the diagrams.  In the first approach the exact
calculation of all loop integrals is performed, followed by their
expansion in the external momenta.  In the second approach we used
asymptotic expansion \cite{Smirnov:1990rz} and expanded the diagrams
in external momenta before performing the loop integrals, finding the
same result as with the first approach.  The final expressions
collected in Table~\ref{tab:llgres} are compact and simple.

As mentioned above, if the external particles in the flavor violating
lepton-photon vertex are on-shell, gauge invariance requires
$F_{VL}^{fi}=F_{VR}^{fi}=0$ for $i\neq f$.  As the diagrams involving
dimension-$6$ vertices may have complicated tensor structures, the
vanishing of the $F_{VL}$ and $F_{VR}$ is an important check of our
calculation.  As an additional check we performed the whole
calculation in a general $R_{\xi}$ gauge finding that the $\xi$
dependence cancels for all form-factors.  Here one should keep in mind
that taking into account only 1PI irreducible diagrams is sufficient
for the calculation of $F_{TL}$, $F_{TR}$ - however, taking into
account also lepton, photon and mixing photon-$Z^0$ self-energies
diagrams is obligatory to cancel completely the vector form-factors
and to get a gauge-independent renormalization constant for the
electric charge.

\begin{table}[htb] 
\begin{tabular}{lp{5mm}l}
\hline 
{\small Group (diagrams of Figs.~\ref{fig:llgse},~\ref{fig:llg1pi})}
&& {\small Tensor form-factors} \\[1mm]
\hline\\[-4mm]
$Z^0$ (3a, 2a($Z^0$)) && $F_{TL}^{Z\;fi} = \dfrac{4e \left[
    \left(C_{\vp \ell}^{(1)fi} + C_{\vp \ell}^{(3)fi} \right) m_f
    (1+s_W^2) - C_{\vp e}^{fi} m_i (\frac{3}{2} - s_W^2)\right]}{
  3(4\pi)^2}$ \\[2mm]
&& $F_{TR}^{Z\;fi} = \dfrac{4e \left[ \left( C_{\vp \ell}^{(1)fi} +
    C_{\vp l}^{(3)fi} \right) m_i (1+s_W^2) - C_{\vp e}^{fi} m_f
    (\frac{3}{2} - s_W^2)\right]}{ 3(4\pi)^2}$ \\[3mm]
\hline\\[-5mm]
$G^0$ (3b, 2a($G^0$)) && $F_{TL}^{G^0\;fi}=0$\\[2mm]
&&$F_{TR}^{G^0\;fi}=0$ \\[2mm]
\hline\\[-5mm]
$W$ (3c,d,e,j,k, 2b($W$)) && $F_{TL}^{W\;fi}= - \dfrac{10 e m_f
  C_{\vp \ell}^{(3)fi} }{ 3 (4\pi)^2}$ \\[2mm]
&&$F_{TR}^{W\;fi}= - \dfrac{10 e m_i C_{\vp \ell}^{(3)fi} }{
  3(4\pi)^2}$ \\[3mm]
\hline\\[-5mm]
$G^\pm$ (3f,g,h, 2b($G^\pm$)) && $F_{TL}^{G^\pm\;fi}=0$\\[2mm]
&&$F_{TR}^{G^\pm\;fi}=0$ \\[2mm]
\hline\\[-5mm]
$WG$ ``bubble'' (3i,l, 2c) && $F_{TL}^{WG\;fi}=0$\\[2mm]
&&$F_{TR}^{WG\;fi}=0$ \\[2mm]
\hline\\[-5mm]
contact $4$-fermion (3m, 2d) && $F_{TL}^{4f\;fi} = -
\dfrac{16e}{3(4\pi)^2} \sum_{j=1}^3 C_{\ell equ}^{(3)fijj\star}
m_{u_j} \left(\Delta - \log\dfrac{m_{u_j}^2}{\mu^2}\right)$\\[4mm]
&&$F_{TR}^{4f\;fi} = - \dfrac{16e}{3(4\pi)^2} \sum_{j=1}^3 C_{\ell
  equ}^{(3)fijj} m_{u_j} \left(\Delta - \log \dfrac{m_{u_j}^2}{\mu^2}
\right)$\\[4mm]
\hline\\[-5mm]
contact $4$-lepton (3n, 2e) && $F_{TL}^{4\ell\;fi} = \dfrac{2
  e}{(4\pi)^2} \sum_{j=1}^3 C_{\ell e}^{fjji} m_j$\\[2mm]
&&$F_{TR}^{4\ell\;fi} = \dfrac{2 e}{(4\pi)^2} \sum_{j=1}^3 C_{\ell
  e}^{jifj} m_j$\\[2mm]
\hline
\end{tabular}
\caption{One-loop contributions to form factors $F_{TL}^{fi}$ and
  $F_{TL}^{fi}$ giving rise to $\ell_i\to \ell_f\gamma$ up to order
  $1/\Lambda^2$.\label{tab:llgres}}
\end{table}

We see that in the final result, at the 1-loop level and in the first
order of expansions in $1/\Lambda^2$ and $m_l$, only the five Wilson
coefficients $C_{\ell e}^{fjji}$, $C_{\ell equ}^{(3)fijj}$, $C_{\vp
  \ell}^{(3)fi}$, $C_{\vp e}^{fi}$ and $C_{\vp \ell}^{(1)fi}$ enter,
while the contribution of all other Wilson coefficients is zero.

It is interesting to note that the term proportional to $C^{(3)}_{\ell
  e q u}$ is the only one containing a divergence.  This divergence
must be canceled by a counter-term to $Q_{eW}$ and/or $Q_{eB}$. The
appearance of this divergence can be understood by looking at a UV
complete theory of NP. Consider as an example a theory with a heavy
scalar particle. Directly calculating the contributions to $F_{TL}$
and $F_{TR}$ in the full theory one would obtain a finite
result. However, when matching to full theory on the SM extended with
dimension-6 operators the situation is more complicated: integrating
out the heavy particle at the matching scale $\Lambda$ gives rise to
$C_{eW}$ and $C_{eB}$ at the loop-level and $C^{(3)}_{\ell equ}$ at
the tree-level. However, as all Wilson coefficients, $C_{eW}$ and
$C_{eB}$ can only contain the hard part of the corresponding
loop-contribution while the soft part must be canceled by the
loop-contribution of $Q^{(3)}_{\ell equ}$ to $C_{eW}$ and $C_{eB}$ in
an effective theory. It turns out that the hard part which contributes
to $C_{eW}$ and $C_{eB}$ has a infrared divergence which is canceled
by the UV divergence of the soft part (as can be best seen using
asymptotic expansion).  Comparing this result with the one in the full
theory we see that the $\mu$-dependence in the contribution of
$C^{(3)}_{\ell equ}$ to $F_{TL}$ and $F_{TR}$ must be replaced by the
mass of the heavy scalar, i.e. $\Lambda$.  In our numerical analysis
we neglect (possible but rather exotic in the lepton sector)
contributions from $Q^{(3)}_{\ell equ}$ operator - coefficients of
such lepton-quark contact terms can be independently constrained using
the LHC measurements~\cite{deBlas:2013qqa}.

\subsection{Anomalous magnetic moments and electric dipole moments}

The form-factors listed in Table~\ref{tab:llgres} for $f=i$ can
directly be used to calculate also the electric dipole moments of
charged leptons and the contribution (in addition to the SM) to their
anomalous magnetic moments:
\begin{equation}
 d_{\ell_i} = \dfrac{-1}{\Lambda^2}\, {\rm Im}\left[ {F^{ii}_{TR} }
   \right] \,,
\end{equation}
\be a_{\ell_i}= \dfrac{2m_{\ell_{i}}}{e\Lambda^2}\, {\rm Re}\left[
  {F^{ii}_{TR} }\right] \, .   \ee
The experimental bounds on the EDM of charged leptons are given in
Table~\ref{tab:EDMs}.

The anomalous magnetic moment of the electron is usually used to
determine the fine structure constant, but determining $\alpha_{em}$
from rubidium atom experiments \cite{Cadoret:2008st}, one can still
use it for obtaining bounds on NP
\cite{Girrbach:2009uy,Crivellin:2010ty,Giudice:2012ms}.  For the
anomalous magnetic moment of the muon there is the long known
discrepancy between experiment and the SM prediction for
$a_{\mu}=(g-2)/2$ \cite{Passera:2004bj, Passera:2005mx, Davier:2010nc,
  Hagiwara:2011af, Blum:2013xva}:
\be
\label{AMMbound}
\Delta a_{\mu}\, = \, a^{exp}_{\mu}-a^{SM}_{\mu} \approx \, (2.7 \pm
0.8)\times 10^{-9} \, .
\ee
This discrepancy could point towards physics beyond the SM and, if
verified, could make the search for $\ell_i\to\ell_f\gamma$ decay even
more promising, as both processes depend on the operators with
formally the same field and Dirac structure, differing only by the
choice of flavor indices.

The current experimental limit on the anomalous magnetic moment of the
tau lepton is rather weak, but it can be improved in the future
\cite{Fael:2013ij}:
\begin{equation}
	-0.052\leq a_\tau \leq 0.013\,.
\end{equation}

\newsection{$\ell_i\to \ell_j \ell_k\bar \ell_l$ decay
  rate\label{sec:llll}}

LFV operators of dimension-$6$ also give contributions to another set
of experimentally strongly constrained decays, namely decays of heavy
charged lepton into three lighter charged
leptons\footnote{Experimental bounds are usually given on positively
  charged muon decays, as they do not form bound state with atoms what
  would decrease the accuracy of measurements
  \cite{Jamieson:2006cf}.}.  Such decays can be generated already at
the tree-level by $Z^0$ and neutral Goldstone boson exchange, flavor
violating photon couplings generated by $Q_{eW}$ and $Q_{eB}$
operators, or even directly by the $4-$lepton operators.  In this
Section we list the general expressions for the lowest order
contributions to all such 3-body charged lepton decays.  Since all
operators enter already at the tree-level we choose not to consider
loop-diagrams for these processes.

We split the expressions for the $\ell_i\to \ell_j \ell_k\bar \ell_l$
decays into 3 groups, depending on composition of the final state
leptons:
\begin{itemize}
\item[(A)] Three leptons of the same flavor: $\mu^\pm \to e^\pm e^+
  e^-$, $\tau^\pm \to e^\pm e^+ e^-$ and $\tau^\pm \to \mu^\pm \mu^+
  \mu^-$.
\item[(B)] Three distinguishable leptons: $\tau^\pm \to e^\pm \mu^+
  \mu^-$ and $\tau^\pm \to \mu^\pm e^+ e^-$.
\item[(C)] Two lepton of the same flavor and charge and one with
  different flavor and opposite charge: $\tau^\pm \to e^\mp \mu^\pm
  \mu^\pm$ and $\tau^\pm \to \mu^\mp e^\pm e^\pm$.
\end{itemize}

\begin{figure}[htb]
\begin{center}
\begin{picture}(160,100)(-20,-40)
\ArrowLine(70,0)(0,50)
\Text(25,40)[l]{$p_k$}
\Text(-15,50)[l]{$l^k$}
\Text(35,10)[l]{$\theta$}
\ArrowLine(70,0)(0,-40)
\Text(25,-35)[l]{$p_l$}
\Text(-15,-40)[l]{$\bar l^l$}
\Text(35,-10)[l]{$\theta'$}
\DashLine(-20,0)(70,0){5}
\DashCArc(70,0)(50,145,-150){5}
\ArrowLine(70,0)(140,0)
\Text(150,0)[c]{$l^j$}
\Text(105,10)[c]{$p_j$}
\GCirc(70,0){15}{1.0}
\Text(70,0)[c]{$p_i$}
\end{picture}
\end{center}
\caption{Kinematics of $\ell_i\to \ell_j\ell_k\bar \ell_l$ decay in
  the CMS frame.
\label{fig:llll}}.
\end{figure}
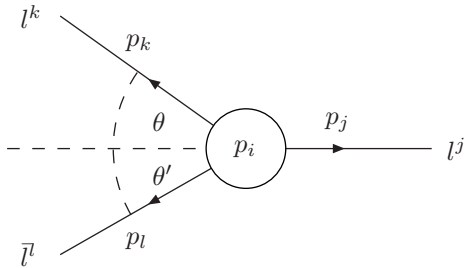

We decompose the amplitude $A$ for the decay $\ell_i\to
\ell_j\ell_k\bar \ell_l$ as
\bea
A = A_0 + A_\gamma\,,
\label{eq:llllampl}
\eea
where $A_0$ contains all operators for which one can neglect the
momenta of the external leptons and $A_\gamma$ is the photon
contribution generated in our approximation by $Q_{eW}$ and $Q_{eB}$
only.  The amplitude $A_0$ can without loss of generality be written
as\footnote{We define the amplitude in such a way that calculating a
  diagram equals $i A$, which means that the Wilson coefficients are
  purely real in the absence of CP violation.}:
\bea
A_0 = \frac{1}{\Lambda^2}\sum_I C_I [\bar u(p_j) Q_I u(p_i)][\bar
  u(p_k) Q^\prime_I v(p_l)]
\label{eq:lllla0}
\eea
with the momenta assignments shown in Fig.~\ref{fig:llll}.  The basis
of quadrilinears $Q_I\times Q^\prime_I$ is given by:
\bea
O_{VXY} & = & \gamma^{\mu}P_X \times \gamma^{\mu}P_Y \,,\nonumber\\
O_{SXY} & = & P_X \times P_Y \,,\nonumber\\
O_{TX} & = & \sigma^{\mu\nu} \times \sigma^{\mu\nu}P_X \,,
\eea
where $X,Y$ stands for the chiralities $L$ and $R$.  For processes
with two identical leptons in the final state one needs to include
crossed diagrams in which the different spinor ordering $[\bar u(p_j)
  Q_I u(p_l)][\bar u(p_k) Q^\prime_I v(p_i)]$ appears.  However, one
can always reduced these contributions to form given in~\eq{eq:lllla0}
by the appropriate Fierz transformations (see
e.g.~\cite{Nieves:2003in}).

The contributions from photon exchange for various types of decays
(A), (B), (C) read (retaining only $1/\Lambda^2$ terms):
\bea
A_\gamma^{(A)} & = & \frac{e v}{\Lambda^2} \left(
\frac{1}{(p_i-p_j)^2} [\bar u(p_j) i\sigma^{\mu\nu} (C_{\gamma L} P_L
  + C_{\gamma R}P_R) (p_i - p_j)_\nu u(p_i)] [\bar u(p_k) \gamma_\mu
  v(p_l)] - (p_j\leftrightarrow p_k) \right) \nonumber\\
A_\gamma^{(B)} & = & \frac{e v}{\Lambda^2}\frac{1}{(p_i-p_j)^2}[ \bar
  u(p_j) i\sigma^{\mu\nu} (C_{\gamma L} P_L + C_{\gamma R} P_R) (p_i -
  p_j)_\nu u(p_i)][\bar u(p_k) \gamma_\mu v(p_l)] \nonumber \\
A_\gamma^{(C)} & = & 0
\label{eq:llllag}
\eea
In~\eq{eq:lllla0} and~\eq{eq:llllag} we did not write explicitly
flavor indices of $C_I, C_\gamma$ but we specify them later in the
next Section.

The general expression for the spin averaged square matrix element
${\cal M} = \frac{1}{2}\sum_{pol}|A|^2$ is complicated, but due to the
hierarchy of the charged lepton masses, in most cases it is sufficient
to assume $m_i\equiv M\gg m_j,m_k,m_l$ and neglect the lighter lepton
masses (which also eliminates the contribution of Goldstone bosons).
Only the contribution from the photon penguin requires more care due
to singularity of photon propagator for small momenta.  For the photon
penguin, in order to get the correct final result one needs to expand
matrix element and phase space kinematics at least up to the order of
$m^2/M^2$.  Then, using standard expressions for 3-particle phase
space one can integrate the matrix element and obtain the branching
ratios (for comparison see~\cite{Ilakovac:2012sh}):
\bea
\mathrm{Br}(\ell_i\to \ell_j \ell_k \bar \ell_l) &=& \frac{N_c
  M^5}{6144\pi^3 \Lambda^4 \Gamma_{\ell_i}} \left(4 \left(|C_{VLL}|^2
+ |C_{VRR}|^2 + |C_{VLR}|^2 + |C_{VRL}|^2\right)\right.\nonumber\\
&+&|C_{SLL}|^2 + |C_{SRR}|^2+ |C_{SLR}|^2+ |C_{SRL}|^2 \nonumber\\
&+& \left.   48 \left(|C_{TL}|^2 + |C_{TR}|^2\right) + X_\gamma\right)
\label{eq:br4l}
\eea
where $N_c=1/2$ if two of the final state leptons are identical,
$N_c=1$ in all other cases and $\Gamma_{\ell_i}$ is the total decay
width of the initial lepton.  The photon penguin contribution reads:
\bea
X_\gamma^{(A)} & =& -\frac{16ev}{M}\mathrm{Re}\left[\left(2 C_{VLL} +
  C_{VLR} - \frac{1}{2} C_{SLR} \right) C_{\gamma R}^\star + \left( 2
  C_{VRR} + C_{VRL} - \frac{1}{2} C_{SRL} \right) C_{\gamma L}^\star
  \right] \nonumber\\
&+& \frac{64 e^2v^2}{M^2} \left(\log\frac{M^2}{m^2} - \frac{11}{4}
\right)(|C_{\gamma L}|^2 + |C_{\gamma R}|^2) \nonumber\\
X_\gamma^{(B)} & =& -\frac{16ev}{M}\mathrm{Re}\left[ \left(C_{VLL} +
  C_{VLR}\right) C_{\gamma R}^\star + \left(C_{VRR} + C_{VRL}\right)
  C_{\gamma L}^\star \right]\nonumber\\
&+& \frac{32 e^2v^2}{M^2} \left(\log\frac{M^2}{m^2} -3\right) (|C_{\gamma
  L}|^2 + |C_{\gamma R}|^2) \nonumber\\
X_\gamma^{(C)} &=& 0 
\label{eq:brg}
\eea

\subsection{Decay $\ell_i\to \ell_j\ell_j\bar \ell_j$}

This option responds to the physical decays $\mu\to 3e$, $\tau\to 3e$
and $\tau\to 3\mu$.  In general, at the tree-level diagrams mediated
by photon, $Z^0$, the neutral Goldstone boson and 4-lepton contact
terms can contribute to the matrix element.  The quantities $C_X$
in~\eq{eq:br4l} can be expressed in terms of Wilson coefficients of
operators in Table~\ref{tab:dim6} as (with $C_\gamma^{ji}$ defined in
\eq{eq:phtree}):
\bea
C_{VLL} &=& 2\left( (2s_W^2-1) \left( C_{\vp \ell}^{(1)ji} + C_{\vp
  \ell}^{(3)ji} \right) + C_{\ell\ell}^{jijj} \right) \nonumber\\
C_{VRR} &=& 2 \left( 2 s_W^2 C_{\vp e}^{ji} +
C_{ee}^{jijj} \right)\nonumber\\
C_{VLR} &=& -\frac{1}{2} C_{SRL} =  2s_W^2 \left( C_{\vp
  \ell}^{(1)ji} + C_{\vp \ell}^{(3)ji} \right) + C_{\ell e}^{jijj}\nonumber\\
C_{VRL} &=& -\frac{1}{2} C_{SLR} =  (2s_W^2-1) C_{\vp e}^{ji} +
C_{\ell e}^{jjji} \nonumber\\
C_{SLL} &=& C_{SRR} =C_{TL} =C_{TR} = 0 \nonumber\\
C_{\gamma L}&=& \sqrt{2}C_\gamma^{ij\star} \nonumber\\
C_{\gamma R}&=&  \sqrt{2}C_\gamma^{ji}
\eea

\subsection{Decay $\ell_i\to \ell_j \ell_k\bar \ell_k$}

\noindent Such a decay can be realized as $\tau^\pm\to e^\pm\mu^+\mu^-
e$ or $\tau^\pm\to \mu^\pm e^+e^-$.  The coefficients $C_X$ read:
\bea
C_{VLL} &=& (2s_W^2 - 1) \left( C_{\vp \ell}^{(1)ji} + C_{\vp
  l}^{(3)ji} \right) + C_{\ell\ell}^{jikk} \nonumber\\
C_{VRR} &=&  2 s_W^2 C_{\vp e}^{ji} + C_{ee}^{jikk}
\nonumber\\
C_{VLR} &=&  2s_W^2 \left( C_{\vp \ell}^{(1)ji} + C_{\vp
  \ell}^{(3)ji} \right) + C_{\ell e}^{jikk} \nonumber\\
C_{VRL} &=&  (2s_W^2-1) C_{\vp e}^{ji} + C_{\ell
  e}^{jkki} \nonumber\\
C_{SLR} &=& -2 C_{\ell e}^{jkki} \nonumber\\
C_{SRL} &=& - 2 C_{\ell e}^{jikk} \nonumber\\
C_{SLL} &=& C_{SRR} =C_{TL} =C_{TR} = 0 \nonumber\\
C_{\gamma L}&=& \sqrt{2}C_\gamma^{ij\star} \nonumber\\
C_{\gamma R}&=&  \sqrt{2}C_\gamma^{ji}
\eea

\subsection{Decay $\ell_i^\pm\to \bar \ell_j^\mp \ell_k^\pm \ell_k^\pm$}

Again, only $\tau$ lepton can decay into such channels, $\tau^\pm\to
e^\mp\mu^\pm\mu^\pm$ or $\tau^\pm\to \mu^\mp e^\mp e^\mp$.  In this
case photon and $Z^0$-mediated diagrams are suppressed by
$1/\Lambda^4$ and only contact $4-$lepton diagram can contribute to
these (rather exotic) process.  The coefficients $C_X$ are given by:
\bea
C_{VLL} &=& 2  C_{\ell\ell}^{kikj} \nonumber\\
C_{VRR} &=& 2  C_{ee}^{kikj} \nonumber\\
C_{VLR} &=& -\frac{1}{2} C_{SRL} =  C_{\ell e}^{kikj}\nonumber\\
C_{VRL} &=& -\frac{1}{2} C_{SLR} = C_{\ell e}^{kjki} \nonumber\\
C_{SLL} &=& C_{SRR} =C_{TL} =C_{TR} = 0 \nonumber\\
C_{\gamma L} &=& C_{\gamma R} = 0 \label{llll_last}
\eea

\newsection{Lepton flavor violating $Z^0$ decays}
\label{sec:zll}

The branching ratio for the lepton flavor violating decays of a $Z^0$
boson $Z^0\to\ell^-_f\ell^+_i$ is given by:
\begin{equation}
\mathrm{Br}\left[ {Z^0 \to \ell_f^\pm \ell _i^\mp } \right] =
\frac{m_Z}{24\pi \Gamma_Z} \left[\f{m_Z^2}{2}\left( \left| C_{fi}^{ZR}
  \right|^2 + \left| C_{fi}^{ZL} \right|^2 \right) + \left| \Gamma
  _{fi}^{ZL} \right|^2 + \left| \Gamma _{fi}^{ZR} \right|^2 \right]\,,
  \label{eq:Zll}
\end{equation}
where $\Gamma_Z\approx2.495$~GeV is the total decay width of the $Z^0$
boson.  We included all tree-level contributions and
\bea
\Gamma^{ZL}_{fi} &=&\dfrac{e}{2 s_W c_W}\left( \dfrac{v^2}{\Lambda^2}
\left( C_{\vp l}^{(1)fi} + C_{\vp l}^{(3)fi} \right) + \left( 1 -
2s_W^2 \right) \delta_{fi} \right)\,,\\
\Gamma^{ZR}_{fi} &=& \dfrac{e}{2s_W c_W}\left( \dfrac{v^2}{\Lambda^2}
C_{\vp e}^{fi} - 2s_W^2 \delta_{fi} \right)\,,\\
C^{ZR}_{fi} &=& C^{ZL\star}_{if} =  -\dfrac{v}{\sqrt{2}\Lambda^2} C_Z^{fi}
\eea
where $C_Z^{fi}$ is defined as 
\bea
C_Z^{fi} = \left(s_W C_{eB}^{fi} + c_W C_{eW}^{fi} \right) \,.
\label{eq:cz}
\eea
The experimental bounds on these decays are given in
Table~\ref{table:Zll}.  Their current sensitivities are not as good as
for the other lepton flavor violating decays but a future linear
collider could significantly improve them \cite{Djouadi:2007ik}.  Note
that theoretical prediction in \eq{eq:Zll} is for the decay
$Z^0\to\ell^-_f\ell^+_i$ or $Z^0\to\ell^+_f\ell^-_i$ while the
experimental values are for the sum
$Z^0\to\ell^-_f\ell^+_i+\ell^-_i\ell^+_f$. Therefore, \eq{eq:Zll} must
be multiplied by a factor of 2 in order to compare it to the
experimental values.

\begin{table}[htdp]
\begin{minipage}{2in}
\centering \vspace{0.8cm}
\renewcommand{\arraystretch}{1.2}
  \begin{tabular}{@{}|c|c|}
\hline
Process & Experimental bound\\
\hline \hline
$\mathrm{Br}\left[Z^0\to\mu^\pm e^\mp\right] $ & ~$1.7 \times 10^{-6}$
~\cite{Abreu:1996mj} \\
\hline
$\mathrm{Br}\left[Z^0\to\tau^\pm e^\mp\right] $ & ~$9.8 \times
10^{-6}$ ~\cite{Abreu:1996mj} \\
\hline 
$\mathrm{Br}\left[Z^0\to\tau^\pm \mu^\mp\right] $ & ~$1.2 \times
10^{-5}$ ~\cite{Abreu:1996mj} \\
\hline \hline
\end{tabular}
\end{minipage} 
\caption{Experimental upper limits (95 \% CL) on the lepton flavor
  violating $Z^0$ decay rates.}
\label{table:Zll}
\end{table} 

\newsection{Numerical Analysis}
\label{sec:numerics}

In the absence of fine-tuning and accidental cancellations the Wilson
coefficients of the flavor changing 4-lepton operators and of the
flavor changing $Z^0$-lepton-lepton vertex are most stringently
constrained by the three-body charged lepton decays, while
$C_{\gamma}^{fi} = c_W C_{eB}^{fi} - s_W C_{eW}^{fi}$ is best
restricted by the radiative lepton decays.  Henceforth, as a first
approximation one can obtain the approximate bounds on
$C_{\gamma}^{fi}$ from the experimental upper limits on ${\rm
  Br}[\ell_i\to\ell_f\gamma]$, assuming that all other Wilson
coefficients are negligible:
\bea
\sqrt{\left| C_\gamma^{12} \right|^2 + \left| C_\gamma^{21} \right|^2}
&\le& 2.45 \times 10^{ - 10} \left( \frac{\Lambda }{1\;\rm{TeV}}
\right)^2\sqrt{\frac{{\rm Br}\left[ \mu \to e\gamma \right]}{5.7 \times
    10^{-13}}} \,,\nn
\label{eq:llgconstraints}
\sqrt{\left| C_\gamma^{13} \right|^2 + \left| C_\gamma^{31} \right|^2}
&\le& 2.35 \times 10^{-6} \left( \frac{\Lambda}{1\;\rm{TeV}} \right)^2
\sqrt{\frac{{\rm Br}\left[ \tau \to e\gamma \right]}{3.3 \times
    10^{-8}}} \,,\\
\sqrt{\left| C_\gamma^{23} \right|^2 + \left| C_\gamma^{32} \right|^2}
&\le& 2.71 \times 10^{-6} \left( \frac{\Lambda}{1\;\rm{TeV}} \right)^2
\sqrt{\frac{{\rm Br}\left[ \tau \to \mu\gamma \right]}{4.4 \times
    10^{-8}}} \,.\nonumber
\eea
Here, the numbers dividing the branching ratios are the current
experimental bounds given in Table~\ref{table:RLFVdecays}.  We see
that the resulting bounds are very strong, of the order of $10^{-10}$
for $\mu\to e$ transitions and of the order of $10^{-6}$ for $\tau\to
\mu,e$ transitions for NP at the TeV scale.  This means that, even
though in a renormalizable theory of NP $C_{\gamma}^{fi}$ can only be
induced at the loop level, an additional suppression mechanism is
needed (especially for $\mu\to e\gamma$) in order the make TeV-scale
NP compatible with experiment.

Knowing that $C_{\gamma}^{fi}$ must be tiny one can set them to zero
in order to constrain other Wilson coefficients using the bounds from
the $\ell_i\to\ell_f\ell_f\bar\ell_f$ decay rates. Here we find (again
normalizing the branching ratios to current limits listed in
Table~\ref{table:llllEXP}):
\bea
C_{\mu eee} &\le& 3.29 \times 10^{-5} \left( \frac{\Lambda}{1\;\rm{TeV}}
\right)^2\sqrt{\frac{{\rm Br}\left[ \mu \to eee \right]}{1 \times
    10^{-12}}} \,,\nn
\label{llll_constraints}
C_{\tau eee} &\le& 1.28 \times 10^{-2} \left(
\frac{\Lambda}{1\;\rm{TeV}} \right)^2 \sqrt{\frac{{\rm Br}\left[ \tau
      \to eee \right]}{2.7 \times 10^{-8}}} \,,\\
C_{\tau \mu \mu \mu } &\le& 1.13 \times 10^{-2} \left(
\frac{\Lambda}{1\;\rm{TeV}} \right)^2\sqrt{\frac{{\rm Br}\left[ \tau
      \to \mu \mu \mu \right]}{2.1 \times 10^{-8}}} \,,\nonumber 
\eea
with $C_{\ell_i\ell_f\ell_f\ell_f}$ given by
\bea 
C_{\ell_i\ell_f\ell_f\ell_f}&=& \left(2\left|\right.  C_{\ell \ell}^{fiff} -
0.54\left( C_{\vp \ell}^{(1)fi} + C_{\vp \ell}^{(3)fi} \right)
\right|^2 + 2\left| C_{ee}^{fiff} + 0.46\;C_{\vp e}^{fi} \right|^2 \nn
&+& \left| C_{\ell e}^{fiff} + 0.46\left( C_{\vp \ell}^{(1)fi} +
C_{\vp \ell}^{(3)fi} \right) \right|^2 + \left| C_{\ell e}^{fffi} -
0.54\;\left.C_{\vp e}^{fi} \right|^2\right)^\frac{1}{2} \,.
\label{C_llll}
\eea
From~\eq{llll_constraints} and~\eq{C_llll} we see that also the Wilson
coefficient of the flavor changing 4-lepton and the
$Z^0$-lepton-lepton vertices must be small for $\Lambda\sim{\cal
  O}(1)$~TeV: of the order of $10^{-5}$ for $\mu\to e$ transitions and
on the order of $10^{-2}$ for $\tau\to \mu$ and $\tau\to e$
transitions.  These constraints are less stringent then the ones
derived from radiative photon decays in~\eq{eq:llgconstraints} but one
should keep in mind that unlike $O_{eB}$ and $O_{eW}$, the other
operators are not necessarily induced at the loop-level but can
already be generated at tree-level.

Also the constraints from $Z^0\to\ell_f^\pm\ell_i^\mp$ can be brought
into a form in which one can directly read off the bounds on the
Wilson coefficients:
\begin{align}
\sqrt{\left| C_{\vp \ell}^{(1)12} + C_{\vp \ell}^{(3)12} \right|^2 +
  \left| C_{\vp e}^{12} \right|^2+ \left| C_Z^{12} \right|^2+ \left|
  C_Z^{21} \right|^2 } &\le 0.06 \left( \frac{\Lambda}{1\;\rm{TeV}}
\right)^2 \sqrt{\dfrac{{\rm Br}\left[ Z^0 \to \mu^\pm e^\mp
      \right]}{1.7\times 10^{-6}}}\,,\nn
\sqrt{\left| C_{\vp \ell}^{(1)13} + C_{\vp \ell}^{(3)13} \right|^2 +
  \left| C_{\vp e}^{13} \right|^2+ \left| C_Z^{13} \right|^2+ \left|
  C_Z^{31} \right|^2 } &\le 0.14 \left( \frac{\Lambda}{1\;\rm{TeV}}
\right)^2 \sqrt{\dfrac{{\rm Br}\left[ Z^0 \to \tau^\pm e^\mp
      \right]}{9.8\times 10^{-6}}}\,,
\label{eq:zll_limits} \\
\sqrt{\left| C_{\vp \ell}^{(1)23} + C_{\vp \ell}^{(3)23} \right|^2 +
  \left| C_{\vp e}^{23} \right|^2+ \left| C_Z^{23} \right|^2+\left|
  C_Z^{32} \right|^2 } &\le 0.16 \left( \frac{\Lambda}{1\;\rm{TeV}}
\right)^2 \sqrt{\dfrac{{\rm Br}\left[ Z^0 \to \tau^\pm \mu^\mp
      \right]}{1.2\times 10^{-5}}}\,.\nonumber
\end{align}
These constraints are less stringent than the ones from
$\ell_i\to\ell_f\ell_f\bar\ell_f$ and $\ell_i\to\ell_f\gamma$ but they
put bounds on the linear combination $C_Z^{fi}$ which is orthogonal to
$C_{\gamma}^{fi}$ (see~\eq{eq:phtree} and~\eq{eq:cz}), so that using
both~\eq{eq:llgconstraints} and~\eq{eq:zll_limits} one can
independently constrain both $C_{eW}^{fi}$ and $C_{eB}^{fi}$.

Finally, one can give similar simplified expressions for the bounds
resulting from the anomalous magnetic moments of charged leptons and
from the EDMs.  Neglecting small lepton mass ratios and taking into
account that some of the Wilson coefficients of the 4-lepton and the
$Z^0$-lepton vertices are real in the flavor conserving case we find
for the EDMs:
\bea
d_e &=& - 2.08 \times 10^{-18}\; \mathop{\rm Im}\nolimits \left[ 2
  \times 10^{-5}\; C_{\ell e}^{3113} + C_\gamma^{11} \right] \left(
\frac{1\;\rm{TeV}}{\Lambda} \right)^2e\,{\rm cm}\,,\nn
d_\mu &=& - 2.08 \times 10^{-18}\; \mathop{\rm Im}\nolimits \left[ 2
  \times 10^{-5}\; C_{\ell e}^{3223} + C_\gamma^{22} \right] \left(
\frac{1\;\rm{TeV}}{\Lambda} \right)^2e\,{\rm cm}\,,\label{EDM_num}\\
d_\tau &=& - 2.08 \times 10^{-18}\; \mathop{\rm Im}\nolimits \left[
  C_\gamma ^{33} \right] \left( \frac{1\;\rm{TeV}}{\Lambda}
\right)^2 e\,{\rm cm}\,,\nonumber
\eea
and for the anomalous magnetic moments:
\bea
a_e &=& 1.17 \times 10^{-6}\; \mathop{\rm Re}\nolimits \left[ 2 \times
  10^{-5} \;C_{\ell e}^{3113} + C_\gamma^{11} \right] \left(
\frac{1\;\rm{TeV}}{\Lambda} \right)^2\,,\nn
a_\mu & =& 2.43 \times 10^{-4}\; \mathop{\rm Re}\nolimits \left[ 2
  \times 10^{-5} \; C_{\ell e}^{3223} + C_\gamma^{22} \right] \left(
\frac{1\;\rm{TeV}}{\Lambda} \right)^2\,,\label{AMM_num}\\
a_\tau & =& 4.1 \times 10^{-3} \; \mathop{\rm Re}\nolimits \left[
  10^{-5}\times\; \left( 1.6\; C_{\vp \ell }^{(1)33} + 2.0\;C_{\ell
    e}^{3333} - 1.7\;\left(C_{\vp \ell}^{(3)33} +C_{\vp e}^{33}\right)
  \right) + C_\gamma^{33} \right]\left( \frac{1\;\rm{TeV}}{\Lambda}
\right)^2\,.\nonumber
\eea
Here we kept the loop induced contributions from the $Q_{\ell e}$ and
$Q_{\vp e}$ since they are not (or weakly) constrained from other
processes.

\begin{figure}
\begin{tabular}{ccc}
\includegraphics[width=0.328\textwidth]{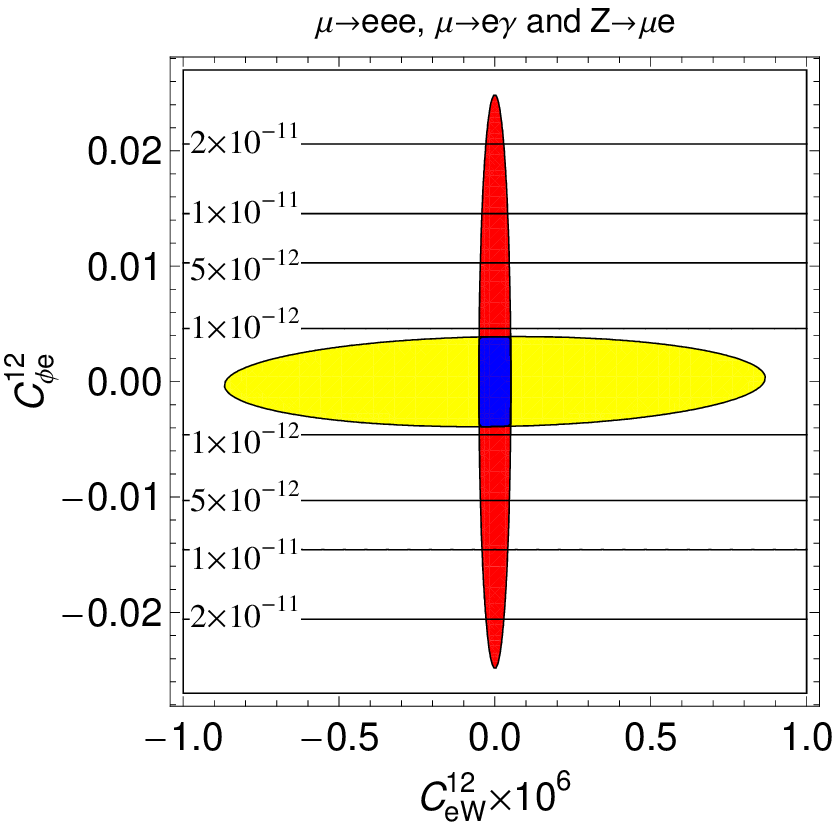} & 
\includegraphics[width=0.31\textwidth]{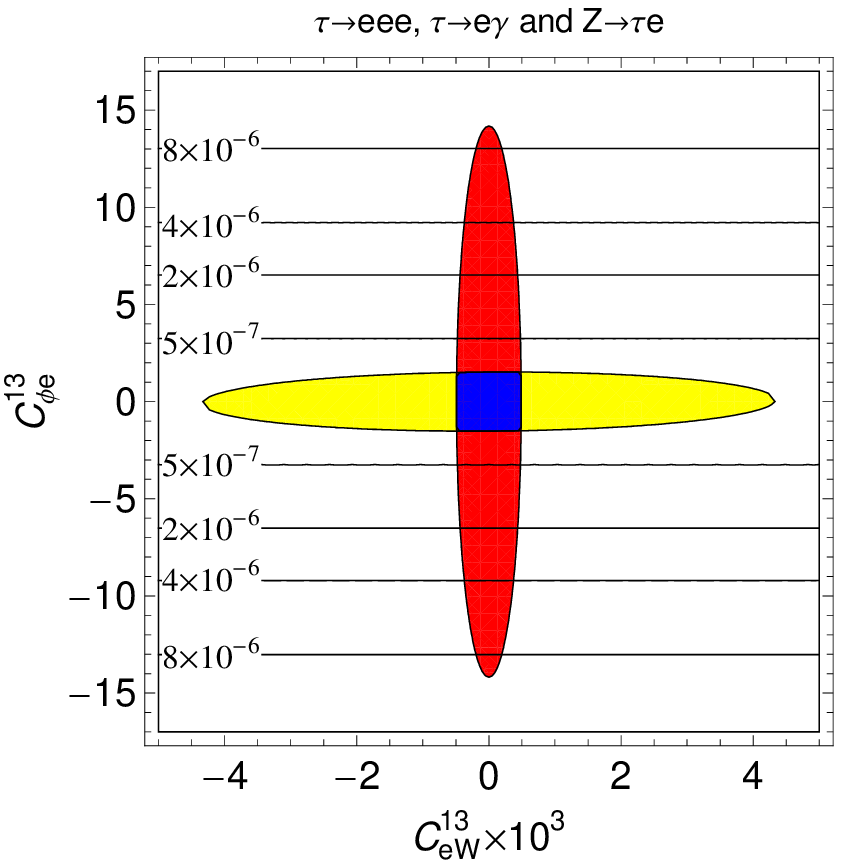} & 
\includegraphics[width=0.31\textwidth]{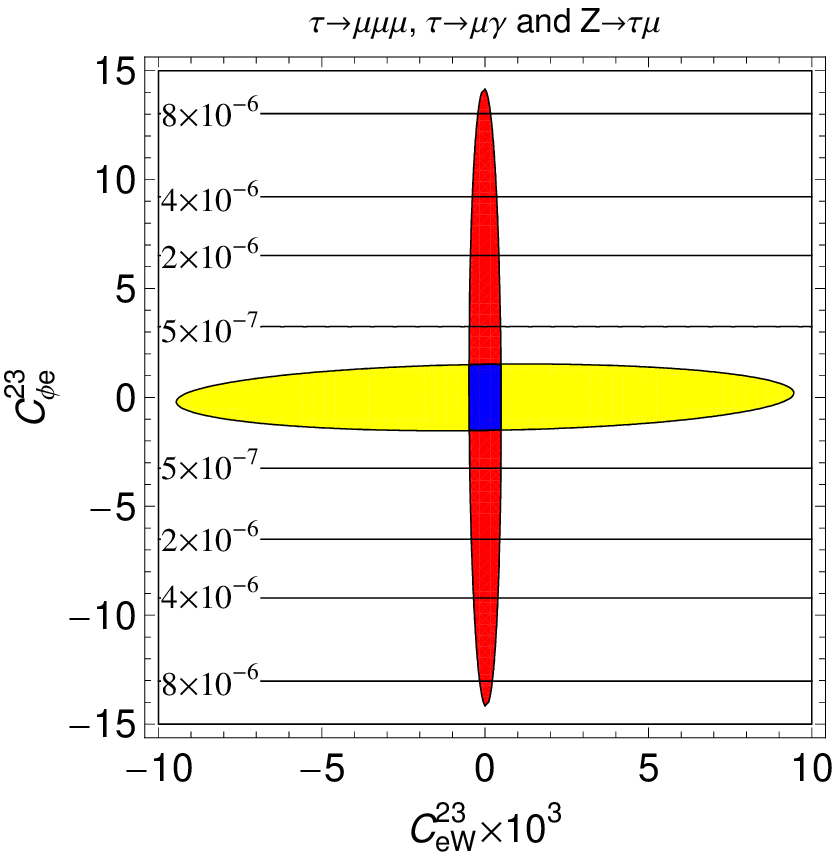}
\end{tabular}
\caption{Allowed regions in the $C_{eW}^{fi}$--$C_{\vp e}^{fi}$ plane
  for $\Lambda=10$~TeV.  Yellow (lightest): $\ell_i\to\ell_f\ell_f\bar\ell_f$,
  red (gray): $\ell_i\to\ell_f\gamma$. The blue region is
  allowed by both decay modes simultaneously. The contour lines show
  the predicted branching ratio for $Z^0\to\ell_f\ell_i$. Note that in
  the parameter space plotted, the dependence of ${\rm
    Br}[Z^0\to\ell_f\ell_i]$ on $C_{eW}^{fi}$ is very weak.}
\label{fig:constraints}
\end{figure}

In order to illustrate the interplay between different Wilson
coefficients in $\ell_i\to\ell_f\gamma$ and
$\ell_i\to\ell_f\ell_f\bar\ell_f$ decays let us consider as an example
the dependence of both decays on the Wilson coefficients of the
operators $O_{\vp e}^{fi}$ and $O_{eW}^{fi}$, as shown in
Fig.~\ref{fig:constraints}.  We see that the regions which respect
both the bound from $\ell_i\to\ell_f\gamma$ and
$\ell_i\to\ell_f\ell_f\bar\ell_f$ are very small, especially for
$\mu\to e$ transitions.  We also show the predicted branching ratios
for $Z^0\to\ell_f\ell_i$ to illustrate that in this plane indirect
limits from the other two processes are currently stronger then the
directly measured upper bounds given in Table~\ref{table:Zll}.

\begin{figure}
\begin{tabular}{ccc}
\includegraphics[width=0.328\textwidth]{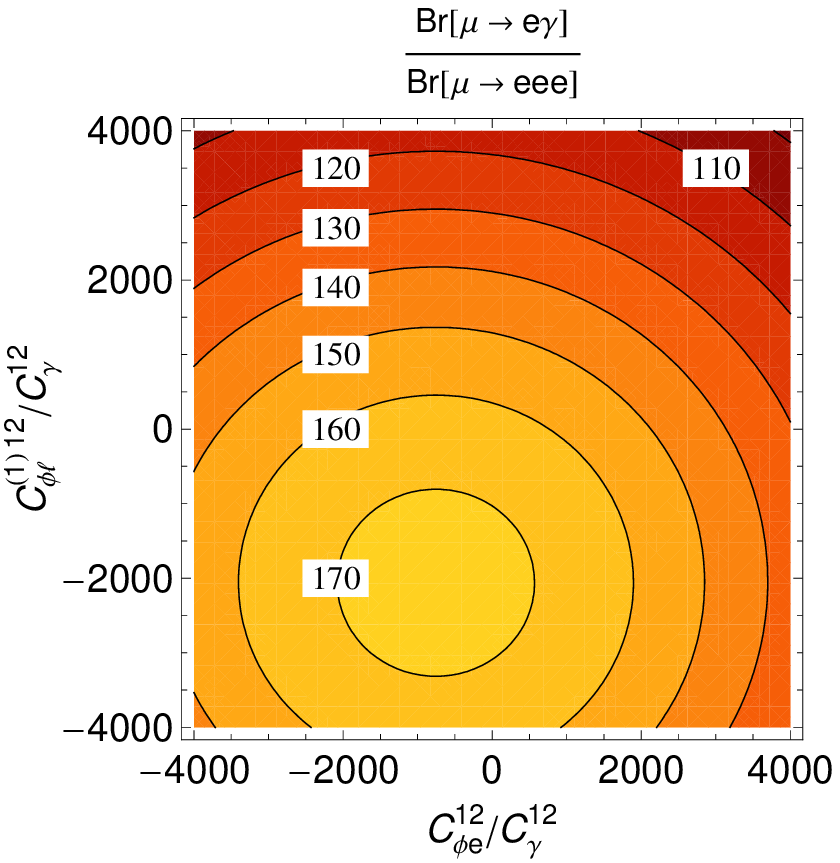}
\includegraphics[width=0.31\textwidth]{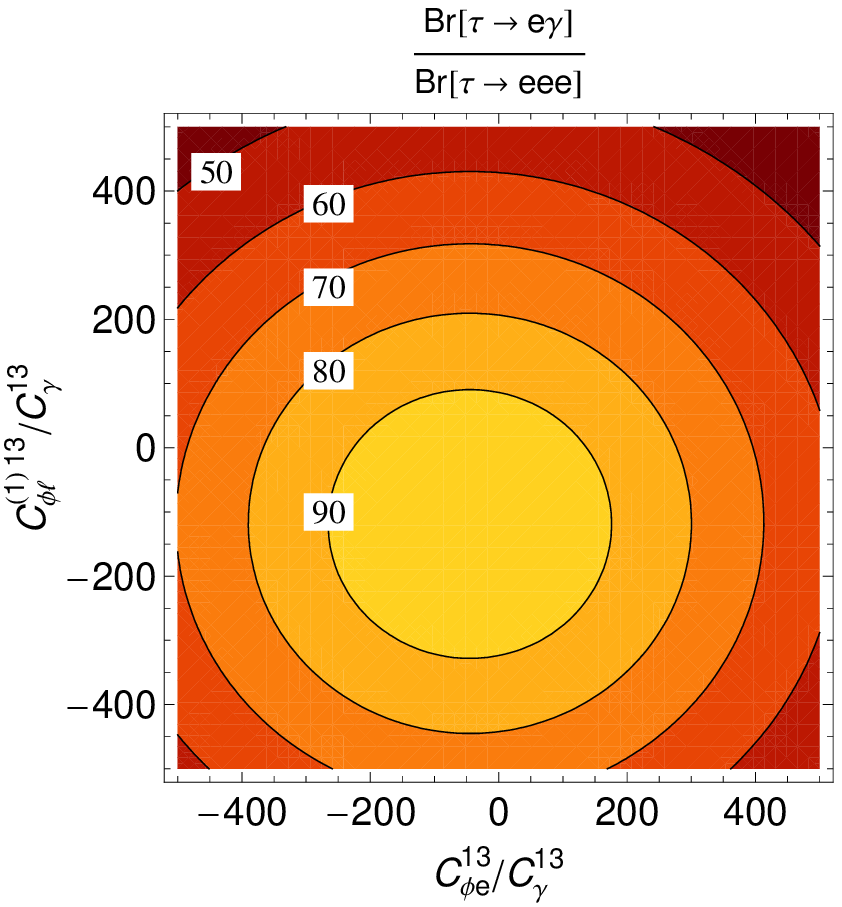}
\includegraphics[width=0.31\textwidth]{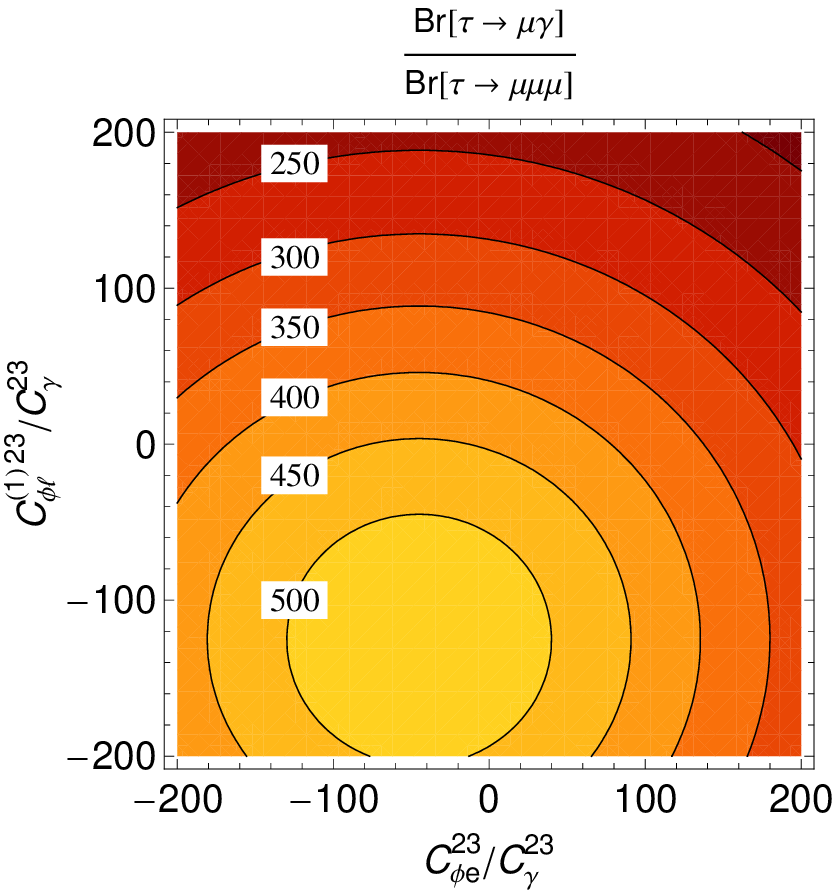}
\end{tabular}
\caption{Ratios ${\rm Br}[\ell_i\to\ell_f\gamma]/{\rm
    Br}[\ell_i\to\ell_f\ell_f\bar\ell_f]$ in the $\frac{C_{\vp
      e}^{fi}}{C_{\gamma}^{fi}}$--$\frac{C_{\vp
      \ell}^{(1)\,fi}}{C_{\gamma}^{fi}}$ plane (independent of the
  scale $\Lambda$ of NP).
\label{fig:ratios}}
\end{figure}

Another interesting aspect is the correlation between the radiative
lepton decays and the three-body charged lepton decays.  In
Fig.~\ref{fig:ratios} we show as an example the ratios ${\rm
  Br}\left[\ell_i\to\ell_f\gamma\right]/{\rm
  Br}\left[\ell_i\to\ell_f\ell_f\bar \ell_f\right]$ as a function of
$\frac{C_{\vp e}^{fi}}{C_{\gamma}^{fi}}$ and $\frac{C_{\vp
    \ell}^{(1)\,fi}}{C_{\gamma}^{fi}}$.  Note that such ratios are
independent of the scale $\Lambda$ of NP and depend only on the ratios
of Wilson coefficients. Thus, given a specific model, one can
determine the branching ratio for one process in terms of the other
one independently of the scale of new physics and also of other
possible cancellations of NP model parameters which can occur in the
ratios $\frac{C_{\vp e}^{fi}}{C_{\gamma}^{fi}}$ and
$\frac{C_{\vp\ell}^{(1)\,fi}}{C_{\gamma}^{fi}}$. As known in the
  literature, the ratio of both decay rates in case in which only $C_{\gamma}^{fi}$ is non-zero depends solely on SM parameters and is given
  by $1/(\frac{\alpha}{3\pi} (\log\frac{m_f^2}{m_i^2}-\frac{11}{4}))$
  (which corresponds to points $(0,0)$ in Fig.~\ref{fig:ratios}).
  From Fig.~\ref{fig:ratios} one can see that contributions from $C_{\vp e}^{fi}$ and
$C_{\vp\ell}^{(1)\,fi}$ can only slightly enhance but more significantly suppress
  this ratio. This is important from the point of view of planned new
  experiments searching for $\mu\to eee$ with increased sensitivity.

\begin{figure}
\begin{tabular}{ccc}
\includegraphics[width=0.32\textwidth]{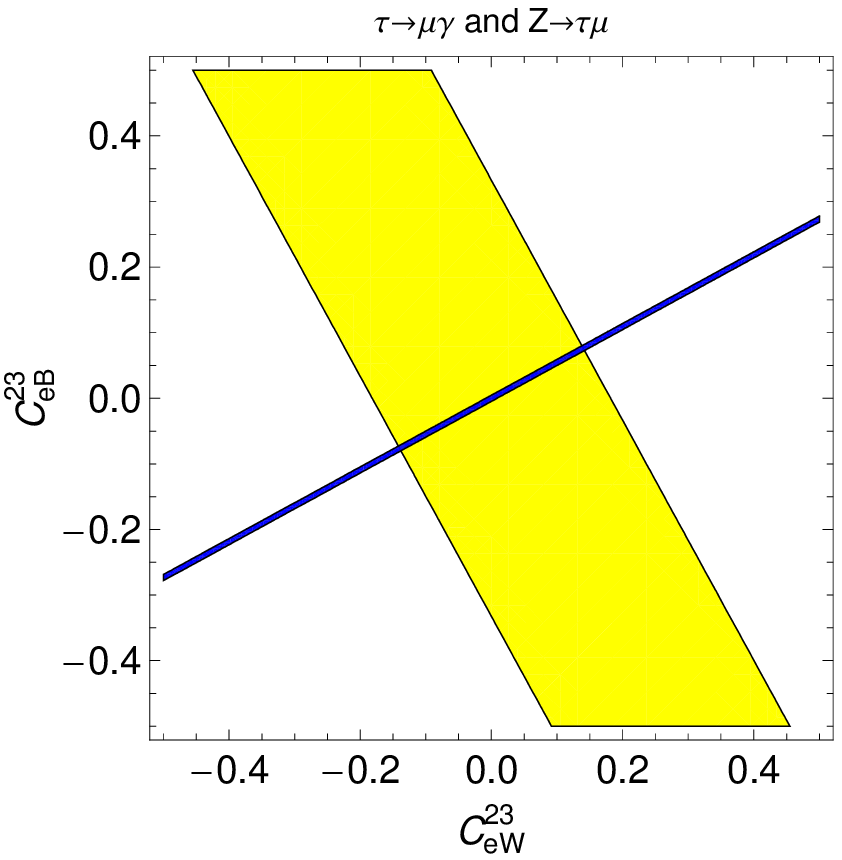}&
\includegraphics[width=0.32\textwidth]{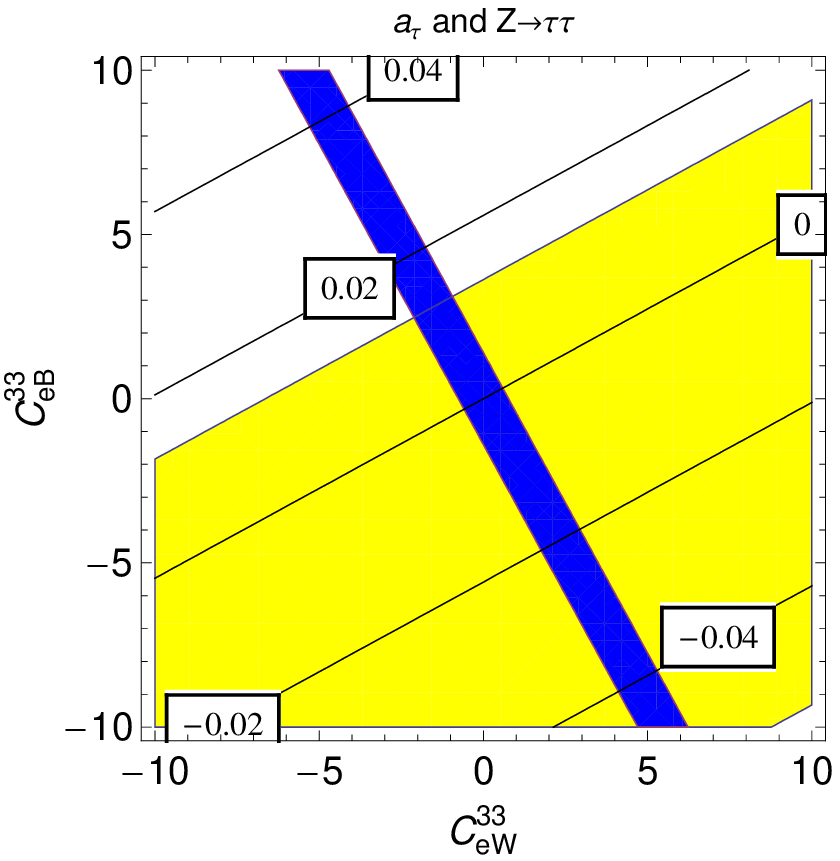} &
\includegraphics[width=0.32\textwidth]{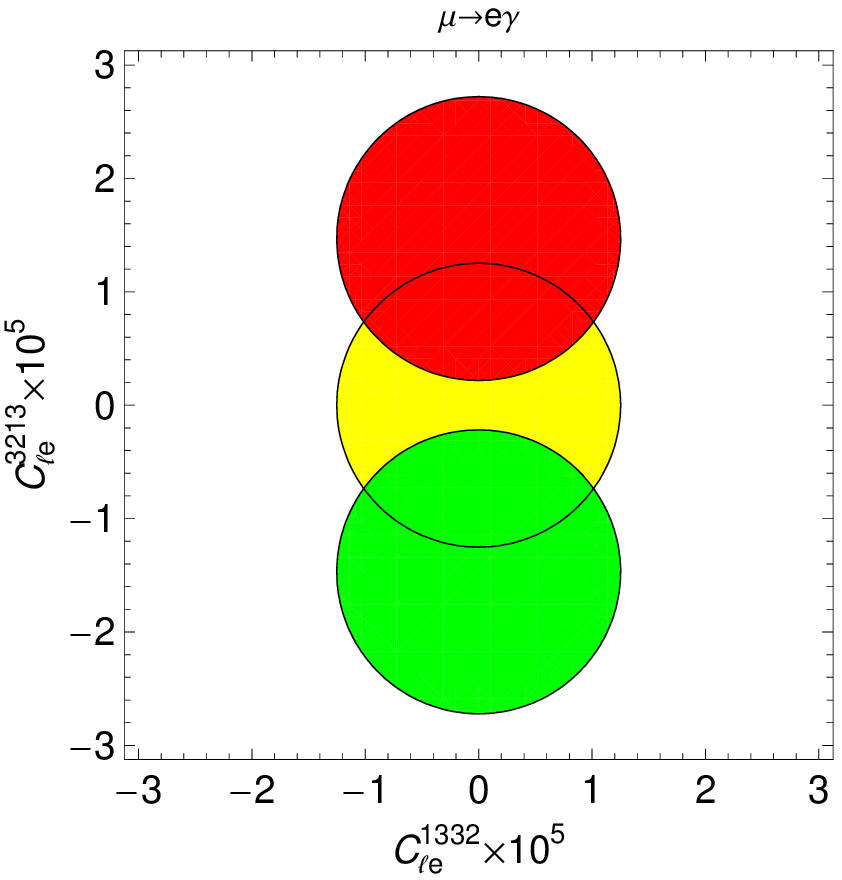}
\end{tabular}
\caption{Left: Allowed regions from Br$[Z^0\to\tau\mu]$ (yellow) and
  Br$[\tau\to\mu\gamma]$ (blue) in the $C_{eW}^{23}$--$C_{eB}^{23}$
  plane for $\Lambda=1$ TeV.
Middle: Correlations between the anomalous magnetic moment of the
$\tau$ lepton and $Z^0\to\tau\tau$.  Yellow (light grey): region
allowed by the $a_\tau$, blue (dark grey): region allowed by
$Z^0\to\tau\tau$.  The contour lines indicate the value of $a_\tau$
for $\Lambda=1$ TeV.
Right: Allowed regions from Br$\left[\mu\to e\gamma \right]$ in the
$C_{\ell e}^{1332}$--$C_{\ell e}^{3213}$ plane for $\Lambda=1$ TeV and
different values of $C_{eW}^{12}$.  Yellow: $C_{eW}^{12}=0$, red:
$C_{eW}^{12}=6\times 10^{-8}$, green: $C_{eW}^{12}=-6\times 10^{-8}$.
\label{Zll_AMM}}
\end{figure}

As observed in Sec.~\ref{sec:zll}, processes involving photon and
$Z^0$ couplings to leptons constrain ``orthogonal'' combinations of
the Wilson coefficients of the operators $O_{eB}$ and $O_{eW}$.  Thus,
using a suitable pair of measurements, one can obtain absolute upper
bounds on each of $C_{eB}$ and $C_{eW}$. An example of such an
exclusion is shown in the left panel of Fig.~\ref{Zll_AMM}: the bound
on the radiative decay $\tau\to\mu\gamma$ strongly correlates the
allowed values for $C_{eB}$ and $C_{eW}$ values to a thin straight
belt, while $Z^0\to\tau\mu$ bound cuts the length of this belt to a
wider but finite compartment.

\psfrag{We}{\tiny{$eW$}}
\psfrag{Be}{\tiny{$eB$}}

Concerning flavor diagonal transitions we can correlate the anomalous
magnetic moments to the corresponding $Z^0\to\ell\ell$ decays. For the
electron and the muon the constraints from the anomalous magnetic
moments are so strong that no sizable effects of NP in $Z^0\to e e$ or
$Z^0\to \mu\mu$ are possible. However, for the tau lepton the
constraints on NP generated terms from $Z^0\to\tau\tau$ and from the
anomalous magnetic moment are not that different. The allowed region
in the $C_{eW}^{33}$--$C_{eB}^{33}$ plane is shown in the middle plot
of Fig.~\ref{Zll_AMM}. In order to obtain these constraint we used
${\rm Br}\left[Z^0\to\tau\tau\right]=(3.370\pm0.008)\%$ \cite{PDG} and
included radiative corrections into our tree-level expression for
$Z^0\to\ell_f\ell_i$, \eq{eq:Zll}, multiplying it by a correction
factor ${\rm Br}\left[Z^0\to\tau\tau\right]_{\rm SM}/{\rm
  Br}\left[Z^0\to\tau\tau\right]_{\rm tree}$ where ${\rm
  Br}\left[Z^0\to\tau\tau\right]_{\rm SM}$ includes radiative
corrections and can be found in Ref.~\cite{PDG}. We also find that the
precision of $Z^0\to\ell_j\bar\ell_j$ decay width measurements limit
the sizes of $C_{\vp \ell}^{(1)\,jj}$, $C_{\vp \ell}^{(3)\,jj}$ and
$C_{\vp e}^{jj}$ Wilson coefficients so stringently that no sizable
effects in the corresponding anomalous magnetic moments are possible
for any lepton flavor.

Another interesting aspect is that one can constrain some of the
4-lepton contact terms by using only the radiative lepton decays. This
is possible because the 4-lepton operator $O_{\ell e}$ affects the
$\ell_i\to\ell_f\gamma$ amplitude at the 1-loop level, as calculated
in Sec.~\ref{sec:llg1}. Once the values of Wilson coefficients
defining the photon coupling $C_\gamma$ are fixed, the bounds on the
4-lepton couplings can be fairly strong - as illustrated in example in
the right panel of Fig.~\ref{Zll_AMM}. There we see that the bounds on
$C_{\ell e}^{1332}$ and $C_{\ell e}^{3213}$ from $\mu\to e\gamma$ for
$\Lambda=1$~TeV are ${\cal O}(10^{-5})$.  Note that these coefficients
(with double $\tau$ flavor index) cannot be constrained from any other
process considered in this article.

\newsection{ Conclusions \label{sec:concl}}

In these article we calculated the expressions for several
theoretically important and experimentally well constrained lepton
flavor violating processes within the Standard Model extended with the
most general set of effective LFV operators of dimension-$6$ invariant
under the SM gauge group.  We computed the complete set of 1-loop
contributions (to the leading order in $m_{\ell}/m_W$) to the
radiative lepton decays $\ell_i\to\ell_f \gamma$ and to the related
electric dipole moments and anomalous magnetic moments of charged
leptons (see Table~\ref{tab:llgres}).  We also obtained the full
expression for the 3-body charged lepton decay rates
$\ell_i\to\ell_j\ell_k\ell_l$ (\eq{eq:br4l}--\eq{llll_last}) and for
the flavor violating $Z^0\to l_f\bar l_i$ decays taking into account
all possible tree-level contributions.

The predictions for all processes are given in terms of Wilson
coefficients of the effective operators, automatically assuring that
the final results are gauge-invariant (which we confirmed explicitly
in our calculation) and that all relevant contributions are
included. The derived expressions allow to obtain model-independent
bounds on the Wilson coefficients of LFV operators, which can be later
easily compared to their values calculated within specific UV complete
extensions of the SM.

To facilitate the comparison, we included in Sec.~\ref{sec:numerics}
approximate numerical formulae directly relating the Wilson
coefficients to current experimental upper bounds on the discussed
processes (\eq{eq:llgconstraints}--\eq{AMM_num}). We show that bounds
on the effective LFV couplings are already very strong if the scale of
NP is low, ${\cal O}(1)$ TeV, and weaken proportionally to the square
of NP scale. We also illustrated possible correlations between Wilson
coefficients of various dimension-$6$ operators and showed that the
loop contributions to $\ell_i\to\ell_f\gamma$ decays are capable to
constrain 4-lepton operators which would be unbounded otherwise.

\vskip 7mm
\centerline{\bf Acknowledgments}
\vskip 2mm

\noindent The authors would like to thank Stefan Pokorski for
suggesting the subject of this article and helpful
discussions. A.C. is grateful to Christoph Greub for his help and
proofreading the article. We also thank Marco Giovanni Pruna and
Adrian Signer for useful discussions.  A.C. is supported by the Swiss
National Science Foundation (SNF).  The work of S.N. has been suported
by the Marie Curie Initial Training Network of the EU Seventh
Framework Programme under contract number PITN-GA-2009-237920-UNILHC.
She thanks the CERN Theory Division for the hospitality during the
stay there.  The work of J.R. is supported by the National Science
Center in Poland under the research grants DEC-2011/01/M/ST2/02466 and
DEC-2012/05/B/ST2/02597.

\vskip 1cm

\setcounter{equation}{0}
\def\theequation{\thesubsection.\arabic{equation}}

\section*{APPENDIX}

We summarize below the Feynman rules arising from the dimension-$6$
operators after the electroweak symmetry breaking.  $i, i_1, i_2$ and
$f, f_1, f_2$ denote the flavor indices of incoming and outgoing
leptons, respectively.  We list only the vertices actually used in our
tree level or 1-loop calculations. For completeness we also include
few necessary purely SM couplings.

\subsection{Feynman rules involving gauge and Goldstone bosons}.   

\input{\axopath feyrul.axo}

\subsection{Feynman rules for 4-fermion operators}

\input{\axopath feyrul_4l.axo}

\vskip 15mm

\subsection{Vector and scalar form-factors contributing to 
off-shell $\ell_i\to \ell_f\gamma^*$ amplitude.}

Gauge invariance requires that $F_{VL}$ and $F_{VR}$ (``vector'')
form-factors vanish for the on-shell external particles.  Thus,
expressions for them must be proportional to the momentum of the
outgoing photon and they do not contribute to $\ell_i\to\ell_f\gamma$
decay rate.   The ``scalar'' form-factors $F_{SL}$ and $F_{SR}$ does
not need to vanish on-shell, but they also cancel out from this
amplitude after contracting with the photon polarization vector.
Still, those form-factors can enter the expressions for the more
complicated processes.  Thus, we list them below, again splitted into
groups of contributions within which the vector form-factors vanish in
the on-shell limit.  Note that some of them are infinite and require
renormalization.

We give only expressions for left scalar form-factor $F_{SL}$ - the
right one can be obtained from $F_{SL}$ by changing the sign and
exchanging the external fermion masses, i.e.:
\bea
F_{SR} = - F_{SL} (m_i \leftrightarrow m_f)
\eea

\begin{itemize}
\item[$Z^0$] group - diagrams 3a, 2a($Z^0$):
\bea
F_{VL}^{Z\;fi} &=& \dfrac{2 e (1-2s_W^2) Q^2}{ 9(4\pi)^2} \left(
C_{\vp \ell}^{(1)fi} + C_{\vp l}^{(3)fi} \right) \left(1 - 6
\log\frac{m_i m_f}{M_Z^2} \right) \nn
F_{VR}^{Z\;fi} &=& -\dfrac{4 e s_W^2 Q^2}{ 9(4\pi)^2} C_{\vp e}^{fi}
\left(1 - 6 \log\frac{m_i m_f}{M_Z^2} \right) \\
F_{SL}^{Z\;fi} &=& \dfrac{2e}{ 9(4\pi)^2} \left[ m_f (1-2s_W^2) \left(
  C_{\vp \ell}^{(1)fi} + C_{\vp l}^{(3)fi} \right) + 2 m_i s_W^2
  C_{\vp e}^{fi} \right]\left(1 - 6 \log\frac{m_i m_f}{M_Z^2} \right)
\nonumber
\eea
\item[$WG$] group - diagrams 3c,d,e,i,j,k,l, 2b($W$),c and
  photon-Goldstone boson self-energy:
\bea
F_{VL}^{WG\;fi} &=& -\dfrac{2e Q^2}{ 9(4\pi)^2}\left[16 C_{\vp
  l}^{(3)fi} + 6 c_W^2 \left(C_{\vp \ell}^{(1)fi} + C_{\vp
  l}^{(3)fi} \right) \right.   \nn
&+&\left.   3 c_W^2 \left(15 C_{\vp \ell}^{(1)fi} +16 C_{\vp
  l}^{(3)fi} \right) \left(\Delta -
\log\frac{M_W^2}{\mu^2}\right)\right] \nn
F_{VR}^{WG\;fi} &=& - \dfrac{2 e c_W^2 Q^2}{3(4\pi)^2} C_{\vp e}^{fi}
\left[2+ 15 \left(\Delta - \log\frac{M_W^2}{\mu^2} \right) \right] \\
F_{SL}^{WG\;fi} &=& \dfrac{e}{9(4\pi)^2} \left[12 c_W^2 (m_i C_{\vp
    e}^{fi} - m_f (C_{\vp \ell}^{(1)fi} + C_{\vp l}^{(3)fi}) ) - 32
  m_f C_{\vp l}^{(3)fi} \right.  \nn
&+&\left. 3 \left( 15 c_W^2 ( m_i C_{\vp e}^{fi} - m_f (C_{\vp
    \ell}^{(1)fi} + C_{\vp l}^{(3)fi}) ) - 2 m_f C_{\vp l}^{(3)fi}
  \right) \left(\Delta - \log\frac{M_W^2}{\mu^2} \right) \right]
\nonumber
\eea
\item[$G^0$] group - diagrams 3b, 2a($G^0$):
\bea F_{VL}^{G^0\;fi} &=& F_{VR}^{G^0\;fi} = F_{SL}^{G^0\;fi} =
F_{SR}^{G^0\;fi} = 0
\eea
\item[$G^\pm$] group - diagrams 3f,g,h, 2b($G^\pm$):
\bea
F_{VL}^{G^\pm\;fi} &=& F_{VR}^{G^\pm\;fi} = F_{SL}^{G^\pm\;fi} =
F_{SR}^{G^\pm\;fi} = 0
\eea
\item[$4l$] group - contact 4-lepton diagrams 3n, 2e:
\bea
F_{VL}^{4\ell\;fi} &=& -\dfrac{2 e Q^2}{3(4\pi)^2} \sum_{j=1}^3
\left(2 C_{\ell \ell}^{fijj} + C_{\ell e}^{fijj} \left(\Delta - \log
\dfrac{m_{\ell_j}^2}{\mu^2} \right)\right) \nn
F_{VR}^{4\ell\;fi} &=& -\dfrac{2 e Q^2}{3(4\pi)^2} \sum_{j=1}^3
\left(2 C_{ee}^{fijj} + C_{\ell e}^{jjfi} \left(\Delta - \log
\dfrac{m_{\ell_j}^2}{\mu^2} \right)\right) \\
F_{SL}^{4\ell\;fi} &=& -\dfrac{2 e}{3(4\pi)^2} \sum_{j=1}^3 \left(2
C_{\ell \ell}^{fijj} m_f - 2 C_{ee}^{fijj} m_i - (C_{\ell e}^{jjfi}
m_i - C_{\ell e}^{fijj} m_f)\left(\Delta - \log \dfrac{m_{\ell_j}^2}{\mu^2}
\right)\right) \nonumber
\eea
\item[$4f$] group - contact $4$-lepton and $2$-lepton-$2$-quark
  diagrams 3m, 2d:
\bea 
F_{VL}^{4f\;fi} &=& \dfrac{4e Q^2}{9(4\pi)^2} \sum_{j=1}^3 \left(
C_{\ell q}^{(1)fijj} - C_{\ell q}^{(3)fijj} + C_{\ell u}^{fijj}
\right) \left(\Delta - \log\dfrac{m_{u_j}^2}{\mu^2}\right) \nn
&-& \dfrac{2e Q^2}{9(4\pi)^2} \sum_{j=1}^3 \left( C_{\ell q}^{(1)fijj}
+ C_{\ell q}^{(3)fijj} + C_{\ell d}^{fijj} \right) \left(\Delta -
\log\dfrac{m_{d_j}^2}{\mu^2}\right) \nn
F_{VR}^{4f\;fi} &=& \dfrac{4e Q^2}{9(4\pi)^2} \sum_{j=1}^3 \left(
C_{eq}^{fijj} + C_{eu}^{(3)fijj}\right) \left(\Delta -
\log\dfrac{m_{u_j}^2}{\mu^2}\right) \\
&-& \dfrac{2e Q^2}{9(4\pi)^2} \sum_{j=1}^3 \left( C_{eq}^{fijj} +
C_{ed}^{(3)fijj}\right) \left(\Delta -
\log\dfrac{m_{d_j}^2}{\mu^2}\right) \nn
F_{SL}^{4f\;fi} &=& \dfrac{4e}{9(4\pi)^2} \sum_{j=1}^3 \left( m_f
\left( C_{\ell q}^{(1)fijj} - C_{\ell q}^{(3)fijj} + C_{\ell
  u}^{fijj}\right) - m_i \left( C_{eq}^{fijj} +
C_{eu}^{(3)fijj}\right) \right) \left(\Delta -
\log\dfrac{m_{u_j}^2}{\mu^2}\right) \nn
&-& \dfrac{2e}{9(4\pi)^2} \sum_{j=1}^3 \left( m_f \left( C_{\ell
  q}^{(1)fijj} + C_{\ell q}^{(3)fijj} + C_{\ell d}^{fijj}\right) - m_i
\left( C_{eq}^{fijj} + C_{ed}^{(3)fijj}\right) \right) \left(\Delta -
\log\dfrac{m_{d_j}^2}{\mu^2} \right) \nonumber
\eea
\end{itemize}

\bibliographystyle{hieeetr}
\bibliography{cnr}
\end{document}